\documentclass[%
 reprint,
 amsmath,amssymb,
 aps,
]{revtex4-2}

\usepackage{graphicx}
\usepackage{dcolumn}
\usepackage{bm}

\usepackage{epsfig}
\usepackage{epsf}
\usepackage{mathtools}
\usepackage{indentfirst}
\usepackage{amsmath}
\usepackage{CJK}
\usepackage{amsfonts}
\usepackage{subfig}
\usepackage{rotating}
\usepackage{graphicx}
\usepackage{multirow}   
\usepackage{ragged2e}
\usepackage{graphicx}
\usepackage{comment}
\usepackage[separate-uncertainty,retain-explicit-plus, per-mode = symbol, detect-all=true, range-units=brackets, tight-spacing=true]{siunitx}
\usepackage[dvipsnames]{xcolor} 
\usepackage{float}
\usepackage[version=4]{mhchem}
\usepackage{caption}
\newcommand{\isotope}[2]{\ce{^{#2}#1}}
\newcommand{\ar}[1]{\isotope{Ar}{#1}}

\begin{document}

\preprint{APS/123-QED}

\title{Subsurface cosmogenic and radiogenic production of \isotope{Ar}{42}}  

\author{Sagar S. Poudel}
\email{sagar.sharmapoudel@pnnl.gov}
\author{Ben Loer}
\author{Richard Saldanha}
\author{Brianne R. Hackett}
\author{Henning O. Back}
\affiliation{%
 Pacific Northwest National Laboratory, Richland, Washington, 99354, USA
}%


\begin{abstract}
Radioactive decays from \isotope{Ar}{42} and its progeny \isotope{K}{42} are potential background sources in large-scale liquid-argon-based neutrino and dark matter experiments. In the atmosphere, \isotope{Ar}{42} is produced primarily by cosmogenic activation on \isotope{Ar}{40}. 
The use of low radioactivity argon from cosmogenically shielded underground sources can expand the reach and sensitivity of liquid-argon-based rare event searches. We estimate \isotope{Ar}{42} production underground by nuclear reactions induced by natural radioactivity and cosmic-ray muon-induced interactions. 
 At 3,000 mwe, \isotope{Ar}{42} production rate is $1.8 \times 10^{-3}$ atoms per ton of crust per year, 7 orders of magnitude smaller than the \isotope{Ar}{39} production rate at a similar depth in the crust.
By comparing the calculated production rate of \isotope{Ar}{42} to that of \isotope{Ar}{39} for which the concentration has been measured in an underground gas sample, we estimate the activity of \isotope{Ar}{42} in gas extracted from 3,000 mwe depth to be less than 2 decays per ton of argon per year.
\end{abstract}

\maketitle


\section{Introduction}
\label{sec:intro}
Liquid argon is commonly used as a detection medium for ionizing radiation. It has a high scintillation and ionization yield and allows for the propagation of the scintillation photons and ionization electrons over large distances, making it an ideal target for large neutrino detectors \cite{microboone_acciarri2017design,captain_aguilar2022first, dune2020deep, ICARUS_tortorici2019icarus,DBA_ashitkov2003liquid}, direct-detection dark matter experiments \cite{warp_acciarri2011warp, darkside_UAr_DS50_39Ar, darkside_aalseth2018darkside,amaudruz2018first}, and active scintillation vetos \cite{darkside_aalseth2018darkside, legend_abgrall2021legend}.

Argon is the third-most abundant gas in the Earth's atmosphere, comprising roughly 0.93\% of the atmosphere by volume. Atmospheric argon consists primarily of the stable isotopes \isotope{Ar}{40} (99.6 $\%$), \isotope{Ar}{36},  and \isotope{Ar}{38}. However, due to interactions of high-energy particles produced by cosmic-ray interactions, atmospheric argon also contains three long-lived radioactive isotopes:  \isotope{Ar}{37}, \isotope{Ar}{39}, and \isotope{Ar}{42}. 

\isotope{Ar}{37} decays purely through electron capture and is relatively short-lived (T$_{1/2} \sim 35$ days\cite{nndc}), quickly decaying below measurable levels after the argon is taken underground for use in detectors shielded from cosmic rays.  \isotope{Ar}{39} is a pure $\beta$-emitter with an endpoint energy of \SI{565}{\keV} and a half-life of 268 years. Atmospheric argon contains \isotope{Ar}{39} at an abundance of $8.2 \times 10^{-16}$, corresponding to a specific activity of \SI{1}{Bq/kg$_{Ar}$}\cite{39Arbenetti2007measurement}, which is a significant source of background for low-energy experiments such as dark matter detectors. To reduce the background from decay of \isotope{Ar}{39}, the next generation of argon-based dark matter detectors propose to use argon extracted from deep underground. While the concentration of \isotope{Ar}{39} in atmospheric argon is maintained by interactions of cosmogenic neutrons and other high-energy particles \cite{Saldanha_2019}, production rates underground are significantly reduced \cite{Mei_underground_PhysRevC.81.055802, vsramek2017subterranean}. The DarkSide-50 collaboration has demonstrated that the underground argon they used as their dark matter target has an \isotope{Ar}{39} rate of $7.3 \times 10^{-4}$ \si{Bq/kg_{Ar}}\cite{darkside_UAr_DS50_39Ar}, a factor $\sim 1400$ below atmospheric levels.  

\isotope{Ar}{42} is a radioactive isotope of argon that undergoes beta decay with a half-life of 32.9 years and endpoint energy (Q$_\beta$) of 599 keV\cite{nndc}. Despite having a similar endpoint energy to \isotope{Ar}{39}, the decay of \isotope{Ar}{42} is typically not a concern as the specific activity in atmospheric argon is on the order of \SI{100}{\micro\becquerel/kg$_{Ar}$} \cite{barabash2016concentration, ackermann2013gerda, DEAP_ajaj2019electromagnetic}, four orders of magnitude lower than that of \isotope{Ar}{39}. However, \isotope{Ar}{42} decays to \isotope{K}{42}, whose energetic decay can be a concern, especially in  liquid argon-based neutrino experiments. \isotope{K}{42} ($T_{1/2}$= 12 h) has two major decay modes: i) direct beta decay ($Q_{\beta}$= 3525 keV, BR=81 $\%$); and ii) beta decay ($Q_{\beta}$= 2001 keV) to an excited state of \isotope{Ca}{42} followed by a prompt 1524 keV gamma emission from the \isotope{Ca}{42} \cite{nndc}.  GERDA, an experiment searching for the neutrinoless double beta decay of \isotope{Ge}{76} at 2039 keV, used an array of germanium detectors surrounded by a liquid argon veto. The energetic betas and gammas resulting from the \isotope{K}{42} decay in the argon were a critical background, leading the GERDA collaboration to launch an extensive \isotope{K}{42} background measurement and mitigation campaign \cite{GERDA_lubashevskiy2018mitigation}. Further, in a large-scale liquid argon detector like the DUNE far detector \cite{dune2020deep}, \isotope{K}{42} decay can cause pileup and event reconstruction issues.
In addition, \isotope{K}{42} decay would limit the MeV-scale physics reach of 
DUNE, particularly to solar neutrinos and supernova core-collapse neutrinos \cite{church2020dark,bezerra2023large} .

\isotope{Ar}{42} is produced in the atmosphere primarily by cosmogenic activation of \isotope{Ar}{40}. The dominant production channel is through interactions of energetic alpha particles with \isotope{Ar}{40}:  \isotope{Ar}{40}($\alpha$,2p)\isotope{Ar}{42} \cite{peurrung1997expected}. This reaction has an energy threshold of \SI{13.7}{\MeV} \cite{nndc} and occurs primarily in the upper atmosphere where cosmic-ray interactions can produce a high flux of energetic $\alpha$'s. \isotope{Ar}{42} can also be produced by a two-step neutron capture process on \isotope{Ar}{40}, but this process is subdominant  due to the short half-life of the intermediate \isotope{Ar}{41} (109 min), which requires very high neutron flux (like that produced in nuclear tests and explosions) for any significant production \cite{barabash1997estimate, peurrung1997expected}. 

As with \isotope{Ar}{39}, next-generation experiments for which \isotope{Ar}{42} is a significant background concern are looking to use argon drawn directly from underground sources, with the assumption that underground argon will have a significantly lower rate of \isotope{Ar}{42} than atmospheric argon \cite{legend_abgrall2021legend,church2020dark, bezerra2023large}. 

\section{Underground Production Mechanisms}
\label{sec:prod_mech}
There has been little study on the nuclear reactions that can produce \isotope{Ar}{42} underground and on the \isotope{Ar}{42} content of underground argon \cite{poudel2020background}. \isotope{Ar}{42} is not produced as the decay product of any naturally occurring primordial isotopes and so we focus our attention on production mechanisms involving particle interactions with nuclei present underground. There are two main sources of energetic particles deep underground: particles produced by cosmic-ray muons as they pass through the upper crust and particles produced by radioactive decay of unstable isotopes in the crust. The production rate of \isotope{Ar}{42} underground depends on the composition of the crustal target, flux of these particles, and the cross-sections for producing \isotope{Ar}{42}.

We estimate the rate of these processes in two ways. For cosmogenic production, we perform a particle transport simulation tracking cosmic-ray muons and the particles produced by the muon interactions through a large volume of crust, and count the number of \ar{42} atoms produced (residual isotope production). We also obtain the secondary particle fluxes produced by the cosmic-ray muon interactions and the radiogenic particle flux in the modeled crust and use TALYS \cite{TALYS_1goriely2008improved}\cite{TALYS_2koning2012modern}-given cross-sections to estimate \ar{42} production.  The latter method is also used to estimate the production rate of other short-lived isotopes such as \ar{41}, and thereby calculate \ar{42} production by two-step channels. 

\isotope{Ar}{42} can also be produced by reactions on 
\isotope{Ar}{40}, in a gas-filled void in the rock, or within the rock itself. In solid rock, the \ar{42} must first diffuse out of the rock grain into a gas pocket before it can be extracted.  Due to the short half-life, a significant fraction of the \ar{42} decays during this process. The diffusion and bulk transport times are difficult to estimate and depend strongly on details of the rock composition and structure, so we do not attempt to estimate them in this work. Instead, we separately calculate and report rates for production in rock (\ar{42} atoms per ton of rock per year) and directly on \ar{40} in gas pockets (\ar{42} atoms per ton of argon gas per year).  To estimate the total \ar{42} decay rate in underground argon, we compare to the \ar{39} rate measured by DarkSide-50. This yields an upper limit since (a) the measured value is likely the result of an atmospheric incursion~\cite{incursion_renshaw1239080procuring} and (b) much more \ar{42} will decay before escaping the solid rock than \ar{39} due to the shorter half-life. 

In Section~\ref{sec:crust} we present the crustal composition assumed for this work. In Section \ref{sec:cosmic_flux} and \ref{sec:rad_flux}, we discuss the cosmic-ray muon-induced and radiogenic particle flux in the Earth's crust.
In Section~\ref{sec:transport}, FLUKA-based particle transport and simulations settings are discussed. In Section~\ref{sec:cosmo_prod} and \ref{sec:radio_prod}, we present our evaluation of cosmogenic and radiogenic \isotope{Ar}{42} production rates respectively. Expected \ar{42}/\ar{40} in the crust and \ar{42} activity in underground argon are discussed in Section \ref{sec:final_activity}.

\section{Crust Composition}
\label{sec:crust}
\begin{figure}[h]
\centering
\includegraphics[width=1.0\linewidth]{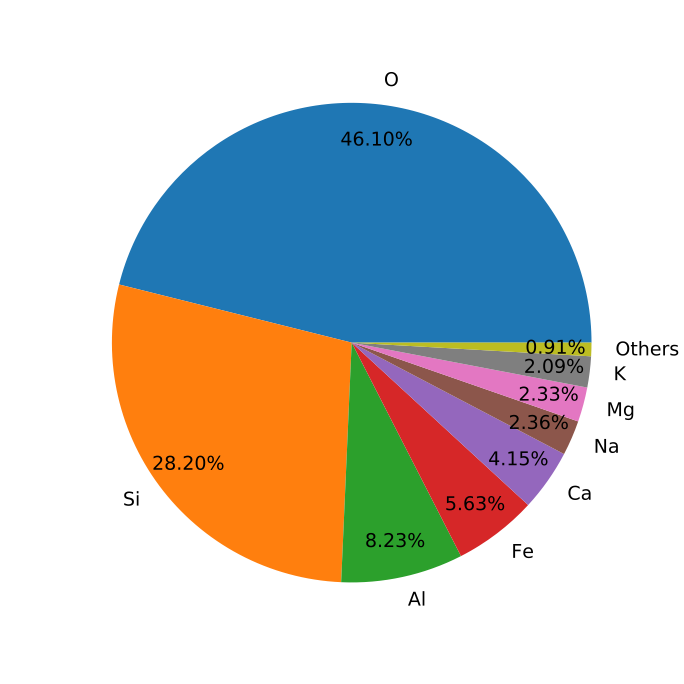}
\caption{Fractional elemental abundances considered for the modeled crust~\cite{rumble2020crc}.} 
\label{Figure: contin_crust}
\end{figure}

We assume a standard continental crust composition with elemental abundances taken from ~\cite{rumble2020crc} and implemented down to 10 ppm level in the simulations. The pie-chart (in Figure~\ref{Figure: contin_crust}) shows distribution of the elemental abundances that are implemented in the modeled rock. Natural isotopic abundances are considered for all elements. The continental crust density is taken as \SI{2.7}{g/cm^3}. 

\section{Cosmogenic Muon Flux}
\label{sec:cosmic_flux}

Cosmic-ray muons can reach great depths in the Earth's crust. As muons propagate through the crustal material, the total flux falls but the mean energy of muons increases as low-energy muons get removed from the spectrum.  The underground muon spectrum spans several orders of magnitude (extending up to thousands of GeV). 
Cosmic-ray muons can produce secondary particles and isotopes primarily through the following processes\cite{malgin2017energy,Zhu_Shirley_2019}: 
\begin{itemize}
    \item Negative muon capture (subdominant beyond 100 meter-water-equivalent (mwe))
    \item Direct muon spallation on nuclei
    \item Muon-induced electromagnetic and hadronic showers.
\end{itemize}

Cosmic-ray muons  are generated for a given depth using MUSIC (MUon SImulation Code)\cite{KUDRYAVTSEV2009339}-given muon energy spectra for standard rock and propagated into our simulated crust.
MUSIC  is a package for muon transport through material.
Muon flux attenuation in a rock primarily depends on the rock thickness, density, and composition.  

For this study, the MUSIC muon flux and the muon energy spectra for a standard rock and for depths of 500 mwe and 3,000 mwe were used as the inputs. The choice of 500 mwe depth was done as it was computationally feasible to transport the surface muons (generated by using EXPACS code \cite{sato2015analytical}) to that depth in our modeled crust, and compare the results with the MUSIC results for standard rock. A comparison of the muon flux and energy spectrum at $\sim$ 500 mwe obtained by transport of the EXPACS-generated muons through our modeled crust and the one obtained by propagating MUSIC-given muons at 500 mwe depth (for standard rock) is shown in Appendix B. The choice of 3,000 mwe depth was made so the results could be compared to data that are available at a similar depth.  

Using the muon flux associated with standard rock introduces some systematics in the cosmogenic particle flux and the \isotope{Ar}{42} production rates. Major systematics, including that from muon flux normalization, are briefly discussed
in Section \ref{sec: uncertainties}.
\\ 

\section{Radiogenic Activity}
\label{sec:rad_flux}
Alphas and neutrons produced by radioactive decays in the natural uranium and thorium decay chains can produce radioactive isotopes through interactions with elements in the Earth's crust. Looking at Table \ref{table:list of reactions} in Appendix A, $\alpha$-induced reactions have energy thresholds $>$ 10 MeV. This is greater than the energy of $\alpha$'s 
originating from the uranium and thorium decay chains (maximum energy of 8.9 MeV from the \isotope{Po}{212} $\alpha$ in the thorium decay chain) and therefore \isotope{Ar}{42} production from radiogenic $\alpha$  is not considered.
We have considered spontaneous fission and ($\alpha$,n)-neutrons. The spontaneous fission neutron spectrum was obtained by following the parameterization in \cite{vsramek2017subterranean}. Fission neutrons from \isotope{U}{235} and \isotope{Th}{232} decay chains were not considered since their yield is several orders of magnitude smaller than that from the \isotope{U}{238} decay chain.

\begin{figure}[h]
\centering
\includegraphics[width=1.0\linewidth]{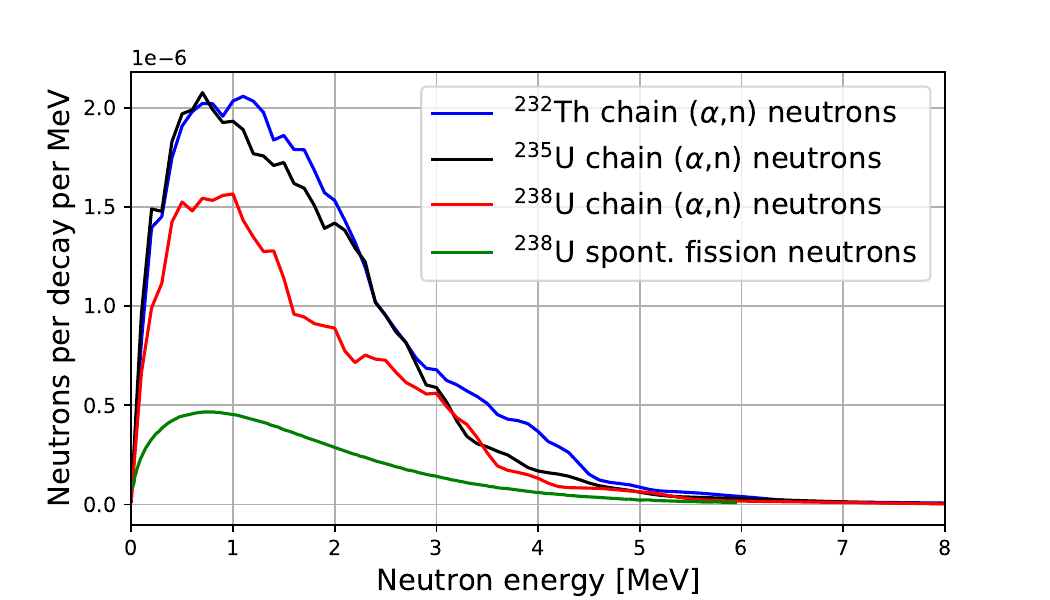}

\caption{Radiogenic neutron yield spectra for a continental crust composition.} 
\label{Figure: neutron_yield}
\end{figure}

We use the NeuCBOT \cite{westerdale2020radiogenic,westerdale_neucBOT} code to obtain ($\alpha$,n) neutron yield and energy spectrum in the crust from uranium and thorium decay chains. 
NeuCBOT uses the SRIM-generated alpha stopping power data for elements \cite{ziegler2004srim,ziegler2010srim}, TALYS ($\alpha$,n) cross-section data,  ENSDF \cite{10.1007/978-3-642-58113-7_227} $\alpha$ decay data, 
and natural isotopic abundance data. With NeuCBOT, we obtain total neutron yield and the energy spectrum of  ($\alpha,n$)  neutrons from \isotope{U}{238}, \isotope{U}{235}, and \isotope{Th}{232} decay chains in the crust. The crustal composition provided as an input to NeuCBOT is the same as the one discussed in Section \ref{sec:crust}, but only includes the most abundant elements that make up 99$\%$ (by mass fraction) of the crust.

The energy spectrum of the ($\alpha,n$) neutrons and the spontaneous fission neutrons are shown in Figure~\ref{Figure: neutron_yield}. The differential neutron yield (on the y-axis) is expressed per decay of the parent isotope in the respective decay chains, assuming secular equilibrium between the $\alpha$-emitting isotopes within individual chains. Total radiogenic neutron yields are listed in Table \ref{table: radio_neutron_yield}. The radiogenic neutron yield of ($\alpha,n$) neutrons is greater than that from spontaneous fission neutrons in the crust. This is expected because the Earth's crust has a high abundance of light elements, and energy thresholds for ($\alpha,n$) reactions for light isotopes are relatively small. The largest neutron yields are from $\alpha$ interactions on the light and relatively abundant isotopes \isotope{Al}{27}, \isotope{Na}{23}, \isotope{Si}{29}, \isotope{Si}{30}, \isotope{O}{18}, \isotope{Mg}{26}, and \isotope{Mg}{25}. 

As a point of comparison, the upper continental crust composition discussed in {{\v{S}}r{\'a}mek} et al.'s paper \cite{vsramek2017subterranean} is similar to the crust composition considered in this work.  Taking  the same uranium and thorium content as reported for the upper continental crust, we find the neutron production rate in our continental crust composition is 20$\%$ higher. This is reasonable as the neutron production rate is particularly sensitive to the assumed elemental abundances of the lighter elements, which are slightly different. 

\begin{table}[h!]
 \small
\begin{tabular}{|p{2.2cm}|p{2.0cm}|p{2.5cm}| }
 \hline
 Source  & Reaction & Neutron yield/decay\\  
 \hline
 \isotope{Th}{232}  chain & ($\alpha$,n)  & 5.279 $\times$ $10^{-6}$\\
 \hline
 \isotope{U}{235} chain & ($\alpha$,n) & 4.819 $\times$ $10^{-6}$ \\
 \hline
  \isotope{U}{238} chain & ($\alpha$,n) & 3.524 $\times$ $10^{-6}$ \\
 \hline
 \isotope{U}{238}  & spont. fission & 1.13 $\times$ $10^{-6}$ \\
 \hline
 \end{tabular}
  \caption{Total radiogenic neutron yield per equilibrium parent isotope decay from various decay chains in the crust}.
  \label{table: radio_neutron_yield}
\end{table}

\section{Particle transport}
\label{sec:transport}
The FLUKA particle physics package (INFN-version) \cite{FLUKA_1_battistoni2015overview,FLUKA_2_bohlen2014fluka} is used to simulate particle interaction and transport,  and to record particle fluence and isotope production. 
FLUKA is a multiparticle transport code that can simulate with high accuracy all relevant particle interactions from keV to GeV scale. The physics models in FLUKA are fully integrated into the code, as discussed in \cite{FLUKA_1_battistoni2015overview,FLUKA_3_ballarini2007physics}.  
A user can incorporate material, geometry, and physics models by calling appropriate FLUKA cards in the input file.
FLAIR, a graphical interface for FLUKA, was used in this work to build and edit input files, link user routines, and construct appropriate executables.

The FLUKA simulation settings adopted in this work are briefly described here. The simulations were performed with  PRECISIO(n) settings, that activate most of the physics processes (electromagnetic, hadronic processes and low-energy neutron interactions) relevant to our interest.  Photonuclear interactions were enabled using the PHOTONUC card and detailed treatment of nuclear excitation was enabled through EVAPORAT(ion) and COALESCE(nce) cards. Full transport of light and heavy ions is enabled in the IONTRANS card. Further, delayed reactions and decay products were enabled using RADDECAY. Neutron interactions at higher energies are handled by FLUKA nuclear models. The interaction and transport of $<$ 20 MeV neutrons are handled by FLUKA's dedicated low-energy neutron libraries that use evaluated neutron data files or measurement data if available. Low-energy neutrons in FLUKA are by default transported all the way to eV energies and lower. 
The USRTRACK card recorded differential fluence (as a function of energy) of the secondary particles produced from the cosmic-ray muon-induced interactions in the modeled crust. Neutron fluence was obtained by recording the  neutrons crossing the layers in the simulated rock. 
The RESNUCLEi card was used to record the residual nuclei produced by the cosmic-ray muon-induced interactions in the simulated crust. A modified user routine usrrnc.f also recorded information about the nuclear reaction leading to isotope production.
\section{Cosmogenic Production}
\label{sec:cosmo_prod}
We calculate \isotope{Ar}{42} production rates by recording \isotope{Ar}{42} isotopes produced in the simulated volume of crust using FLUKA's  nuclear models and low-energy neutron cross-section libraries. In addition, with FLUKA's residual isotope recording, we include all the cosmic-ray muon-induced interactions on all the relevant isotopes present in the crust, including direct muon spallation and heavy-ion collision.

As a cross-check, we also calculate the \isotope{Ar}{42} production rate by combining the TALYS \isotope{Ar}{42} production cross-sections ( calculated for selected nuclear reactions) and the cosmogenic secondary particle flux obtained by particle transport in FLUKA. For this estimate, only the nuclear reactions from cosmic-ray produced light secondary particles (neutrons, protons, deuterons, tritons, and alphas) are considered.

\subsection{Muon Propagation}
Secondary particles resulting from muon interactions  are mainly produced in particle showers, primarily in hadronic showers \cite{malgin2017energy,Zhu_Shirley_2019}. 
Our simulations show that 3-6 m of crustal thickness is enough to ensure full development of hadronic showers without significant attenuation of the muon flux. The cosmic-ray muon-induced particle flux is approximately constant within that thickness. For 500 mwe runs, we allow muons to propagate through a larger thickness (15 m) to also account for neutrons from negative muon-capture and direct muon spallation, and then apply muon-flux correction. The crust is modeled as a cuboidal solid of 20 m x 20 m x 6 m (15 m)  dimensions, with a density 2.7 g/cm$^3$, composed of a homogeneous material of elemental isotopes with natural isotopic abundance.   
The muon propagation and particle transport in the modelled crust is done using FLUKA simulations. 
MUSIC-given muon spectra and the muon flux for a standard rock composition were used as inputs. Muon energies were sampled from the energy distributions and propagated into the simulated crust. Only vertical muons were considered, and the total muon flux is assumed to be entirely that of vertical muons. Separate simulations were run for depths 500 mwe and 3,000 mwe using MUSIC-given muon spectra for respective depths. For each depth, separate simulations were run with positive and negative muons. The flux normalization is done assuming a positive-to-negative muon flux ratio of 1.3:1 \cite{muon_charge_ratio_agafonova2010measurement}.


\subsection{Cosmogenic Particle Flux}
The fluence of the cosmic-ray muon-induced particles generated in the muon-induced shower was recorded using FLUKA. The USRTRACK card recorded the particle fluence (counts cm$^{-2}$ GeV$^{-1}$ per incident primary muon) as a function of particle kinetic energy. Using the muon flux at a given depth, the muon-induced particle flux was then obtained.  Only the flux of neutrons, protons, deuterons, tritons, alphas, and \isotope{He}{3} were recorded, since the flux of these particles is highest in the muon-induced showers. USRTRACK cannot be used to estimate the low-energy (keV-scale) neutron fluence unless the energy binning is very fine. Instead, the neutron flux was obtained by recording neutrons on an event-by-event basis across multiple crustal layers using the modified FLUKA user routine mgdraw.f.
With the input muon spectrum and flux at each depth, we obtained the neutron flux and energy spectrum in the Earth's crust. The total cosmogenic neutron flux at 3,000 mwe in continental crust is calculated to be $2.02 \times 10^{-9}$ n/cm$^2$/sec, similar to the one estimated at Gran Sasso ($2.72 \times 10^{-9}$ n/cm$^2$/sec) reported in \cite{Mei_underground_PhysRevC.81.055802}). However, it is important to note that neutron flux strongly depends on the muon flux and spectrum as well as on the composition of the rock. The total cosmogenic neutron flux at depths 500 mwe and 3,000 mwe in the continental crust are given in Table \ref{table:cosmo_neutron}.  

\begin{figure}
\centering
 
\subfloat[]{
	\label{subfigure:a}
	\includegraphics[width=0.5\textwidth]{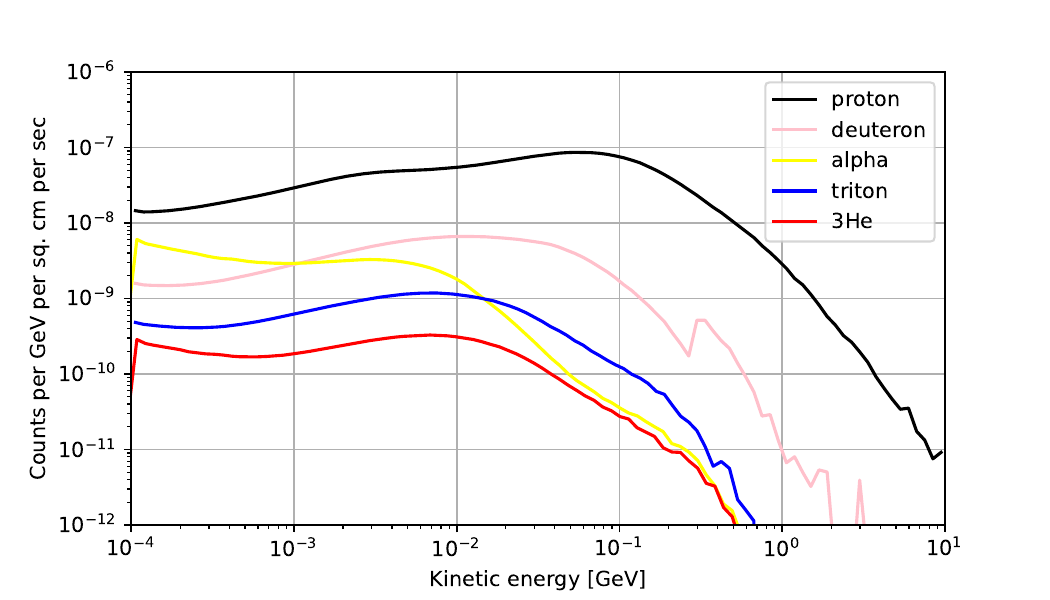} } 
 
\subfloat[]{
	\label{subfigure:b}
	\includegraphics[width=0.5\textwidth]{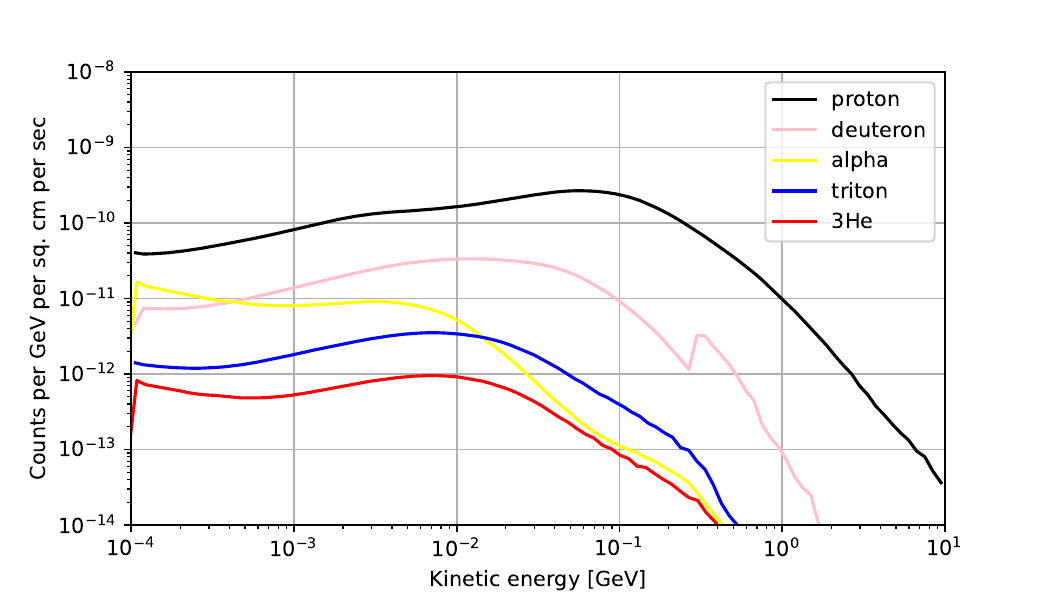}} 
	
\subfloat[]{
	\label{subfig:c}
        \includegraphics[width=0.5\textwidth]{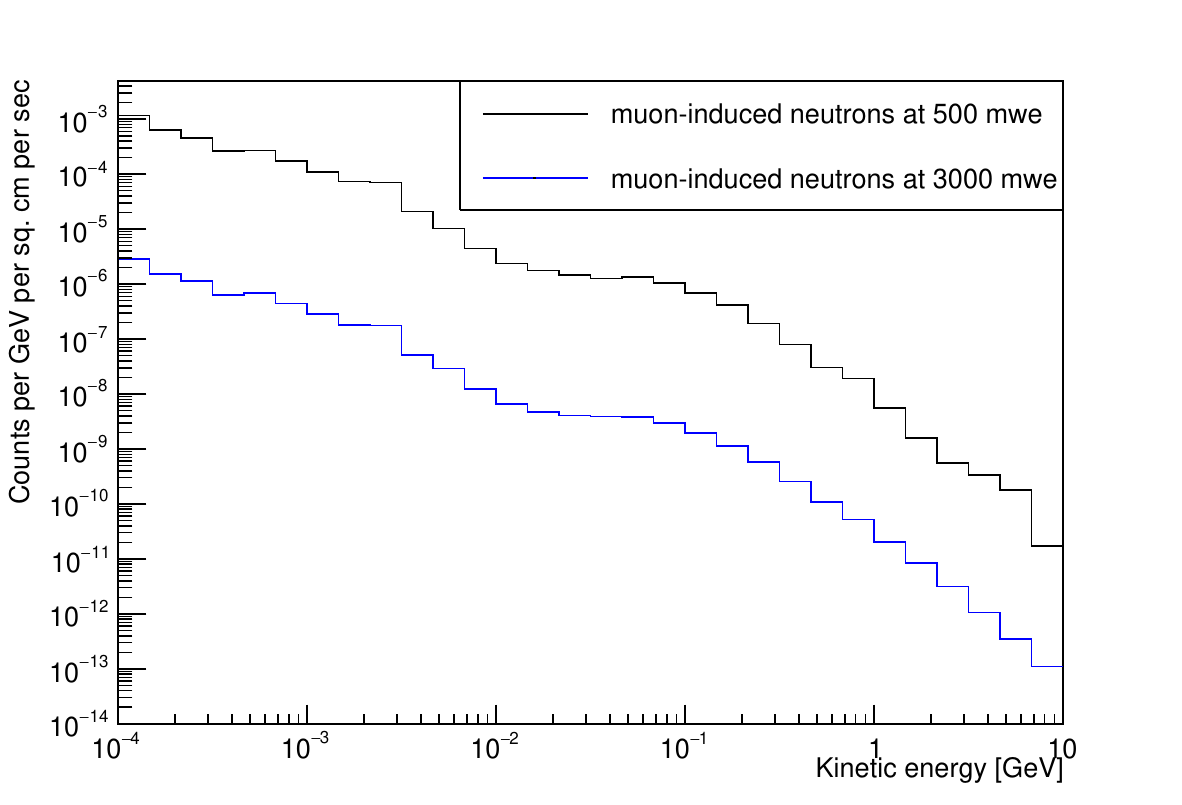}}
 
 \caption{Cosmic-ray muon-induced particle flux at (a) depth 500 mwe and (b) 3,000 mwe for continental crust (obtained using FLUKA USRTRACK). Only the particles with KE $<$ 10 GeV particles were recorded. (c) Muon-induced neutron flux at 500 mwe and 3000 mwe for the same composition (obtained from event-by-event tracking with FLUKA).} 
\label{fig:Cosmo_flux} 
\end{figure}

The cosmogenic particle flux for all particles considered (as a function of kinetic energy) is shown in Figure  \ref{fig:Cosmo_flux}. 

 
\begin{table}[h!]
 \small
\begin{tabular}{|p{1.5cm}|p{2.2cm}|p{3.7cm}|}
 \hline
  Depth & Muon flux\newline (muons/cm$^{2}$/s) & Cosmogenic neutron flux (neutrons/cm$^{2}$/s) in the crust\\  
 \hline
 500 mwe & $2.07 \times 10^{-5}$  & $7.67 \times 10^{-7}$\\
 \hline
 3,000 mwe & $3.09 \times 10^{-8}$ & $2.02 \times 10^{-9}$ \\
 \hline
 \end{tabular}
 \caption{Cosmogenic neutron flux at depths 500 mwe and 3,000 mwe in the crust. Muon flux in the standard rock for corresponding depths are taken from \cite{KUDRYAVTSEV2009339}}
 \label{table:cosmo_neutron} 
\end{table}

\subsection{Cosmogenic Production Rates}
With FLUKA simulations, \isotope{Ar}{42} isotopes were recorded in selected volume of the modeled crust. The RESNUCLEi card was used to obtain the \isotope{Ar}{42} production yield (number of \isotope{Ar}{42} isotopes per cubic cm of crust per primary muon). Alternatively, we also investigated the nuclear reactions, at the individual event-level, that resulted in \isotope{Ar}{42} production using a customized FLUKA user routine usrrnc.f. Production yield given by RESNUCLEi accounts for isotope production by all nuclear reactions including the ones produced by low-energy neutrons ($<$ 20 MeV neutrons in FLUKA) interactions. However, usrrnc.f output does not account for isotope production by low-energy neutron interaction, as information on low-energy neutron-induced residual isotope production is not available in FLUKA at the event level. Muon flux provides the basis for normalization to obtain the production rates from yields obtained by both methods.  

Cosmogenic \isotope{Ar}{42} production rates, obtained by recording the \isotope{Ar}{42} isotopes produced by cosmic-ray muon-induced showers in a volume of simulated crust, are listed in Table \ref{table:cosmo_residual} for depths of 500 mwe and 3,000 mwe. In this case, the FLUKA nuclear models are at work and \isotope{Ar}{42} production from all cosmic-ray muon-induced interactions, including muon spallation and heavy-ion collision, are considered.

\begin{table}[h!]
 \small
\begin{tabular}{l|c|c }
 \hline
 Depth & Muon flux & Cosmogenic \isotope{Ar}{42} produced \\  
 (mwe) & (muons /cm$^2$/s) & (atoms/ton of crust/yr) \\  
 \hline
 500 &  2.07$\times$ 10$^{-5}$ &  0.73 \\
 \hline
 3,000  &  3.09 $\times$ 10$^{-8}$ & 1.8 $\times$ 10$^{-3}$ \\
 
 \hline
 \end{tabular}
 \caption{Cosmogenic  \isotope{Ar}{42} production rates in the crust at depths of 500 mwe and 3,000 mwe. The rates are obtained by recording the \isotope{Ar}{42} isotopes in the modeled crust using FLUKA simulations and normalizing the production yield with respect to the muon flux.}
\label{table:cosmo_residual}
\end{table}

\begin{table}[h!]
 \small
\begin{tabular}{|p{3.0cm}|c|}
\hline
 Nuclear reactions & Contribution to \isotope{Ar}{42} production \\  
 \hline
 \isotope{Ca}{44}(n,3He)\isotope{Ar}{42} &  9 $\%$  \\
  \isotope{Ca}{44}(H*,X)\isotope{Ar}{42} &  12 $\%$  \\
 \isotope{Ca}{44}($\gamma$,X)\isotope{Ar}{42} &  6 $\%$  \\
 \isotope{Ca}{48}(H*,X)\isotope{Ar}{42} &  9 $\%$  \\
 \hline 
 \hline

 \isotope{Fe}{56}(H*,X)\isotope{Ar}{42} &  19 $\%$  \\
 \isotope{Fe}{56}($\pi^-$,X)\isotope{Ar}{42} &  6 $\%$  \\
 \hline
 \hline
Other reactions & 39 $\%$ \\
 \hline
 \end{tabular}
 \caption{Major reactions that produce \isotope{Ar}{42} in the crust.Contribution to the \isotope{Ar}{42} production from various reactions in the modeled crust based on 3,000 mwe simulation runs. The results are based on 32 \isotope{Ar}{42} atoms produced for 1.93 $\times$ 10$^4$ ton-year exposure. As reactions from low energy ($<$ 20 MeV) neutrons were not available on an event-by-event basis, those reactions and their contribution ($\approx$ 10 $\%$ of the total)are not included in the table. In the table, $H^*$ represents heavy ion and $X$ represents a products other than \isotope{Ar}{42}.}
 \label{table: Cosmogenic_production_reaction_type}
\end{table}

The production rates obtained from RESNUCLEi output (given in number of isotopes produced per cubic cm of the crust per primary muon) followed by normalization with respect to MUSIC-given muon flux are shown in the Table \ref{table:cosmo_residual}. \isotope{Ar}{42} production as recorded as an output by usrrnc.f agree within 90$\%$. The agreement is expected as \isotope{Ar}{42} production by $<$ 20 MeV neutrons, which is not recorded by usrrnc.f, is smaller given the high-energy thresholds of direct neutron-induced \isotope{Ar}{42} production.  

At 3,000 mwe depth, the cosmogenic \isotope{Ar}{42} production rate is  $1.8\times 10^{-3}$ atoms per ton of crust per year. The primary channels of \isotope{Ar}{42} argon production and their corresponding rates obtained for 3,000 mwe as obtained from TALYS and FLUKA are shown in Table \ref{table:production-table-final}. Also, combining the statistics from both 500 mwe and 3,000 mwe runs are combined to produce the Table \ref{table: Cosmogenic_production_reaction_type} which  lists the major reactions and their relative contribution to the total cosmogenic \isotope{Ar}{42} production rate. In the table, X represents  products of the nuclear reaction other than \isotope{Ar}{42}. Identity of the heavy ions (represented as $H^*$) was not accessible through the usrrnc.f routine, and no attempt was made to identify the heavy ions through other FLUKA user routines given the high computational cost to do so.

From Table \ref{table: Cosmogenic_production_reaction_type}, one can observe that neutron and heavy-ion induced interactions on isotopes of calcium and iron are dominant contributors to the \isotope{Ar}{42} production in the Earth's crust. The results in the table use the information of all the \isotope{Ar}{42} produced in both positive and negative muon runs for 3,000 mwe depths.

\subsection{TALYS Cross-check of Selected Nuclear Reactions}
We also use the TALYS \isotope{Ar}{42} production cross-sections to estimate \isotope{Ar}{42} production from selected nuclear reactions \cite{TALYS_1goriely2008improved,TALYS_2koning2012modern}.
For a given particle projectile and target isotope, TALYS can give the residual nuclei production cross-section as a function of kinetic energy of the particle projectile. 

The production rate of a particular channel is given by:
\begin{equation}
\label{production_yield}
 P_{i,j} = n_i\int\frac{d\phi_j(E)}{dE}\sigma_{i,j}(E)dE 
\end{equation}
where $P_{i,j}$ is the contribution to the production rate from source $j$ on target isotope $i$, $n_i$ is the number density of target isotope $i$, $\frac{d\phi_j(E)}{dE}$ is the differential (in kinetic energy) flux of particle $j$, and $\sigma_{i,j}$ is the cross-section for the $i(j,X)^{42}\mathrm{Ar}$ reaction. The total production rate is the sum of $P_{i,j}$ over all sources and target isotopes.

All sources (particle projectiles) considered are described in Table \ref{tab:sources} and all the target isotopes considered are presented in Table \ref{table:list of reactions} in Appendix A. 
\begin{table}[h]
    \centering
    \begin{tabular}{c|c}
         Origin & Particle Flux\\
\hline\hline
\multirow{2}{*}{Natural Radioactivity} & Neutron \\ 
 & Alpha \\
\hline
\multirow{5}{*}{Cosmic rays} & Neutron \\
  & Proton \\
  & Deuteron \\
  & Triton \\
  & Alpha \\
\hline
\end{tabular}
\caption{The particle projectiles considered for TALYS-based estimate of \isotope{Ar}{42} argon production. List of the all reactions considered in this case are in Appendix A} 
\label{tab:sources}
\end{table}

Unlike the residual-nuclei recording with FLUKA, TALYS does not simulate the nuclear reactions induced by muons and heavy ions.  Also, TALYS can only simulate nuclear reactions in the energy range of 1 keV to 1 GeV. Energetic cosmic-ray muon interactions can produce secondary particles of energies above 1 GeV. However, the \isotope{Ar}{42} production rate from nuclear reactions induced by particle projectiles with energies $>$ 1 GeV is expected to be small as the cosmogenic particle flux falls quickly at high energies.

The description of the selected reactions and the full list of the reactions considered are given in Table \ref{table:list of reactions} in Appendix A. For all those reactions, the integral in Eq.~\ref{production_yield} was evaluated. 

\begin{figure}
\centering
 
\subfloat[]{
	\label{subfig:a}
	\includegraphics[width=0.45\textwidth]{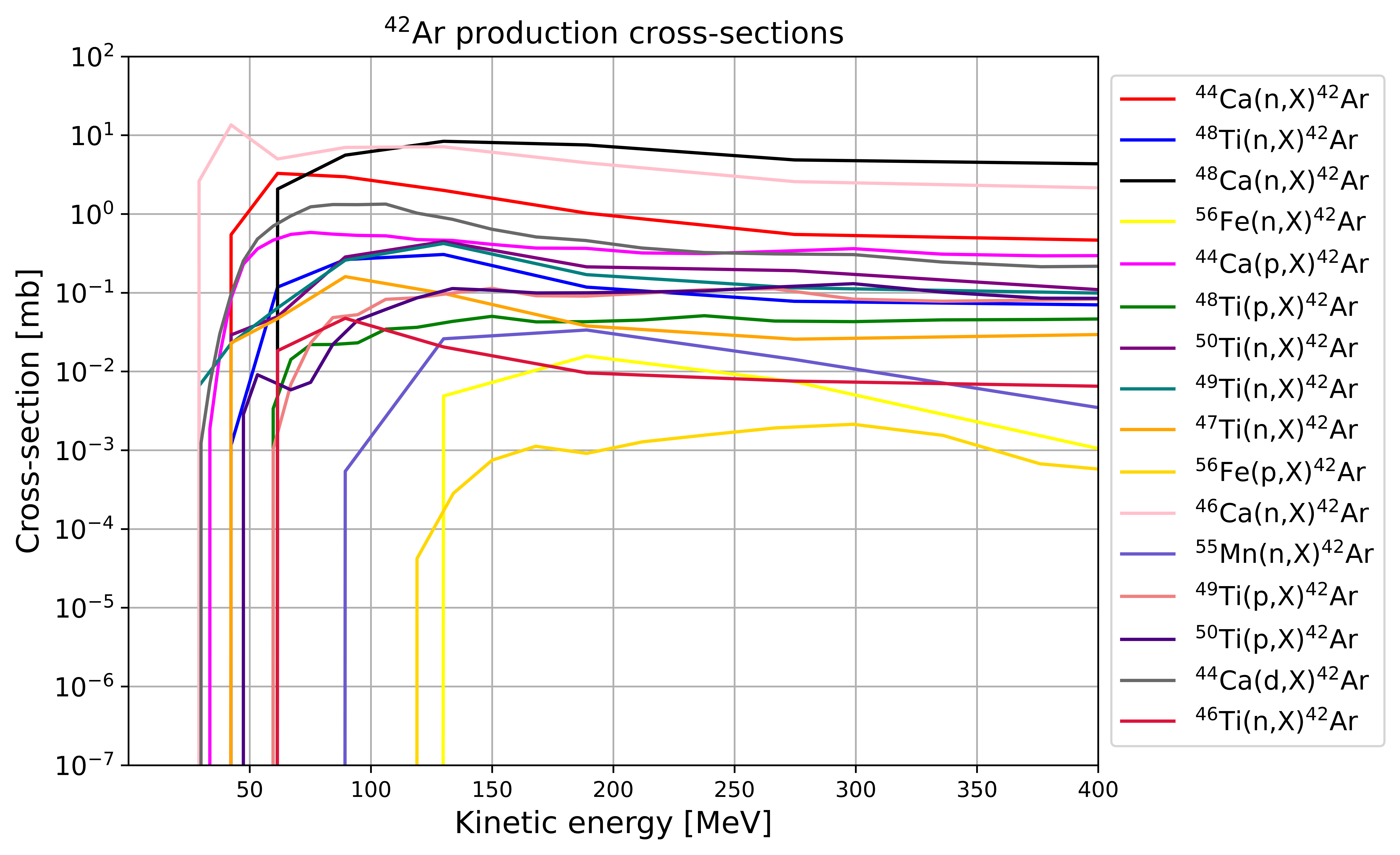} } 
 
\subfloat[]{
	\label{subfig:b}
	\includegraphics[width=0.5\textwidth]{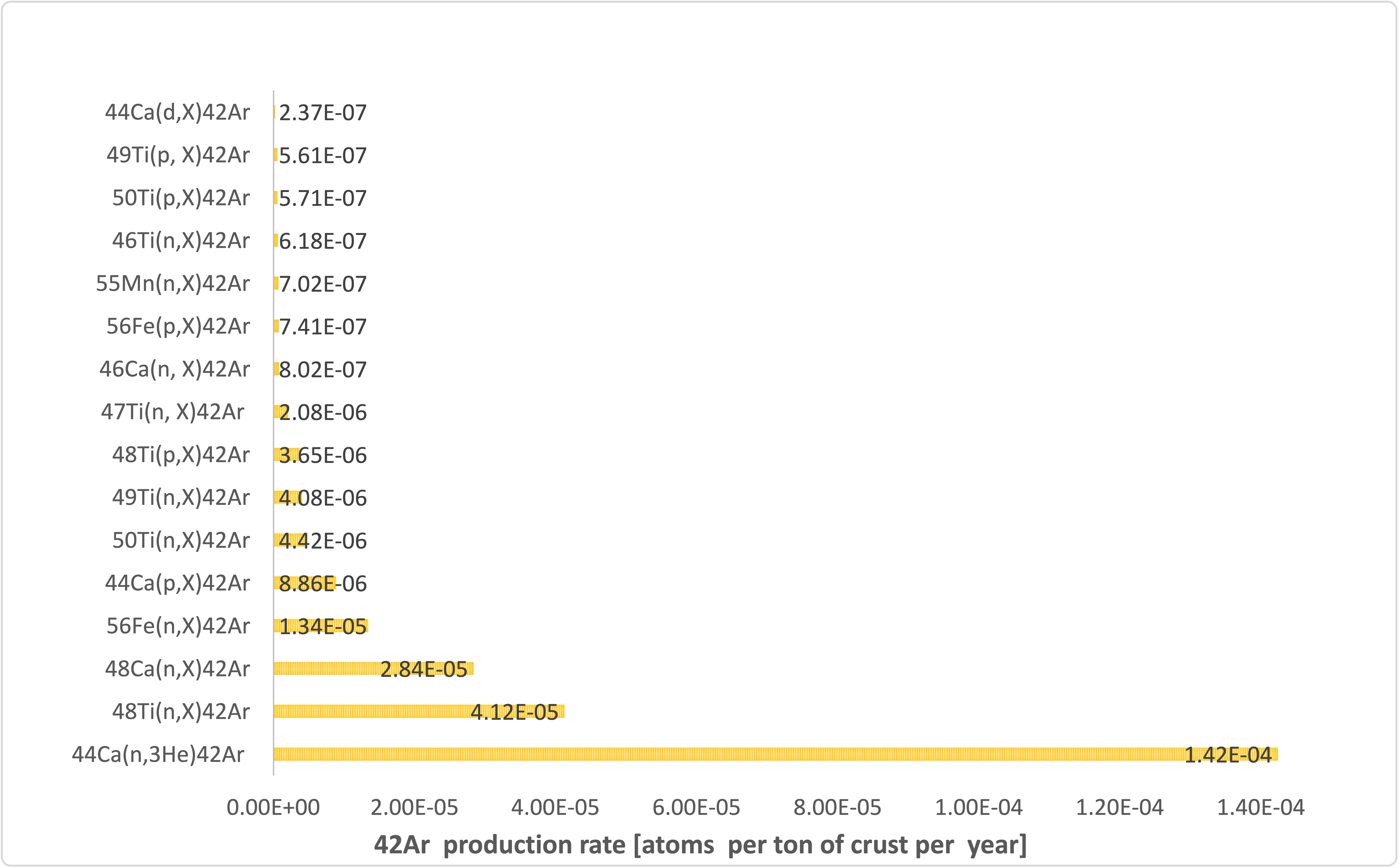} }

 \caption{Major reactions that produce \isotope{Ar}{42} in the crust, and the corresponding production rates obtained using Eq. (\ref{production_yield}) for selected reactions (listed in Table \ref{table:list of reactions} in Appendix A). The TALYS-given production rate at 3,000 mwe is $2.53 \times 10^{-4}$ atoms per ton of crust per year (Note: The FLUKA-based estimate that includes all reactions is an order of magnitude higher). The sum of contributions from all the other considered reactions (not included in this figure) is $\sim$ 1$\%$. (a) the TALYS-given  \isotope{Ar}{42} production cross-sections for those reactions. (b) The production rates from those reactions in the crust at 3,000 mwe. Only n, p, $\alpha$, t, and d induced reactions are included here.}  
 
\label{fig:TALYS_major} 
\end{figure}

The TALYS-based cosmogenic production rate of \isotope{Ar}{42} was calculated for both the 500 mwe and 3,000 mwe depth in the modeled crust. The sum production of \isotope{Ar}{42} from the selected neutron-, proton-, deuteron-, and triton-induced reactions in the crust at 500 mwe and 3,000 mwe are found to be $7.97 \times 10^{-2}$ and $2.53 \times 10^{-4}$ atoms per ton of crust per year, respectively. Figure \ref{fig:TALYS_major} shows nuclear reactions that TALYS identifies as the major contributors to the \isotope{Ar}{42} production in the Earth's crust. Figure \ref{fig:TALYS_major}a shows the cross-sections for those reactions. Figure \ref{fig:TALYS_major}b gives the list of the reactions and their contribution to the \isotope{Ar}{42} production rates at 3,000 mwe. Looking at the cross-sections in Figure \ref{fig:TALYS_major}a and the list of major reactions in Figure \ref{fig:TALYS_major}b, it can be seen that it is primarily due to the high abundance of calcium, titanium, and iron in the crust that the nuclear reactions involving isotopes of these elements dominate. All other reactions not included in Figure~\ref{fig:TALYS_major}b contribute less than 1$\%$ to the total \isotope{Ar}{42} production rate.

The TALYS-based estimate of the cosmogenic \isotope{Ar}{42} production rates are an order of magnitude less than the ones obtained by FLUKA's residual isotope recording. A primary reason for this is , unlike in TALYS, where only a limited set of nuclear reactions were considered, full-fledged simulation with FLUKA includes also the contribution from additional nuclear interactions including heavy-ion collisions and direct muon-spallations.  
Looking selectively at the n, p, $\alpha$, t, and d induced nuclear reactions (shown in Appendix A, Table \ref{table:production-table-final}), the production rates in FLUKA's residual nuclei output are within the same order of magnitude of the TALYS-based production rates (FLUKA's estimate from those reactions are $\sim 70\%$ higher), which is not surprising given the differences in the nuclear models used by those tools. \\ 


\section{Radiogenic Production}
\label{sec:radio_prod}
\begin{figure}[h]
\centering
\includegraphics[width=1.0\linewidth]{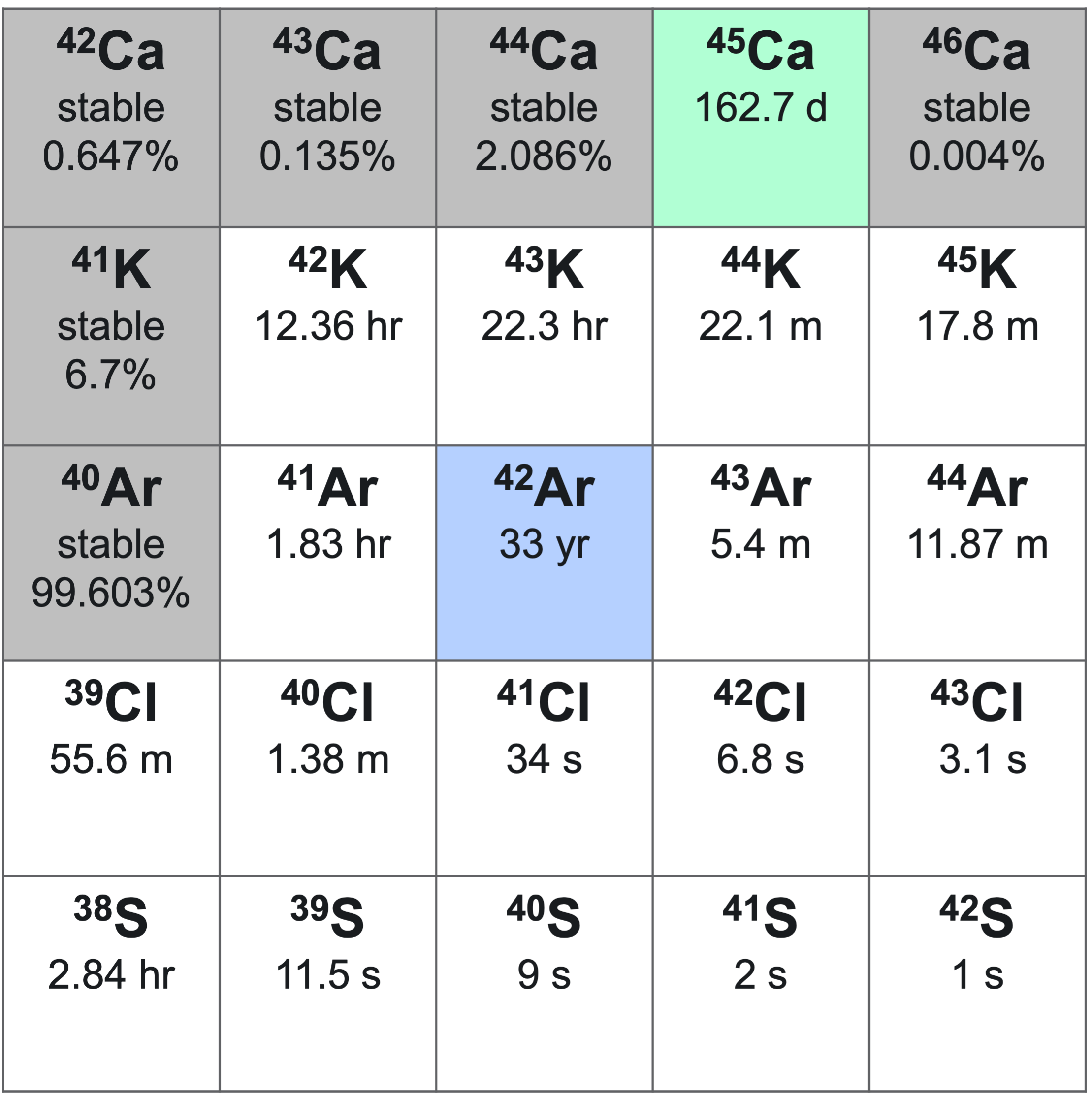}
\caption{Isotopes directly neighboring \isotope{Ar}{42} in mass number table.  The short-lived isotopes (white) directly neighboring \isotope{Ar}{42} (light blue),  one long-lived (light green) and several stable isotopes (dark grey) are shown in the table.}
\label{Figure: mass_number_table}
\end{figure}

Radiogenic production of \isotope{Ar}{42} in the Earth's crust is expected to be significantly suppressed with respect to production in the atmosphere.
Figure \ref{Figure: mass_number_table} shows the isotopes neighboring \isotope{Ar}{42} with respect to mass number. These neighboring isotopes are either short-lived and so are not abundant enough, or are stable so the reactions producing \isotope{Ar}{42} from those isotopes have energy thresholds higher than the energies available to neutrons and $\alpha$'s of radiogenic origin. 

Radiogenic production of \isotope{Ar}{42} in the Earth's crust, however small it may be, could occur through two-step reactions, with intermediate production of the isotopes \isotope{Ar}{41} [$\tau_{1/2}$ = 109 min], \isotope{K}{42} [$\tau_{1/2}$ = 12 hr], and \isotope{Ca}{45} [$\tau_{1/2}$ = 163 d]. Since these isotopes are radioactive and short-lived, and hence trace in concentration, the data on their abundance in the Earth's crust are scarce. We first estimate the cosmogenic and radiogenic production of \isotope{Ar}{41}, \isotope{K}{42}, and  \isotope{Ca}{45}, using the  equilibrium concentration of these isotopes in Eq. (\ref{production_yield}), and then estimate the radiogenic \isotope{Ar}{42} production from neutron-induced reactions on those isotopes.   
\subsection{Radiogenic Neutron Flux}

\begin{figure}[h]
\centering
\includegraphics[width=1.0\linewidth]{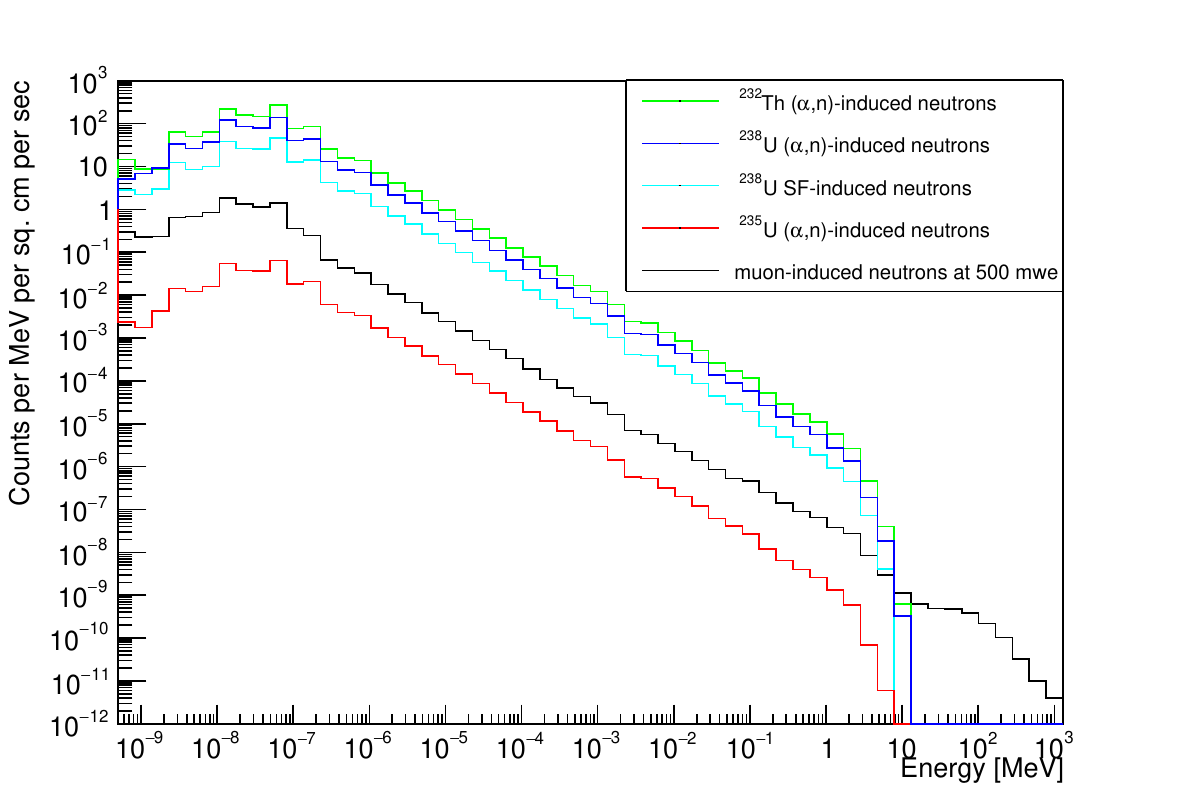}
\caption{Simulated neutron flux and energy spectra from spontaneous fission neutron, ($\alpha$,n) neutrons, and muon-induced neutrons (at 500 mwe) in the Earth's crust.  }
\label{Figure: Neutron_flux}
\end{figure}

We obtain radiogenic neutron flux by propagating the neutrons with energies drawn from the neutron-yield distributions in our modeled crust (shown in Figure \ref{Figure: Neutron_flux}). The neutrons are generated homogeneously and isotropically across a spherical volume of a simulated crust of radius 200 m and density 2.7 g/cm$^{3}$. The elemental composition of the simulated crust is the same as the one discussed in Section \ref{sec:crust}. Separate FLUKA simulations were run by drawing the neutron energies from the neutron yield distributions shown in Figure \ref{Figure: neutron_yield}. FLUKA user routine mgdraw.f tracked the neutrons on an event-by-event basis across multiple rock surfaces. On the surfaces with radii $>$ 50 m, the neutron fluence was constant (within statistical uncertainties). Neutron counts and energies were recorded as neutrons exited one of those large rock surfaces. FLUKA-given neutron fluence (as a function of neutron kinetic energy) was used to obtain radiogenic neutron flux spectra by assuming activity of  uranium (U) and thorium (Th) corresponding to  U:$2.7 \times 10^{-6}$ g/g and Th: $1.05 \times 10^{-5}$ g/g respectively, which are taken from Ref.~\cite{vsramek2017subterranean} as reported for upper continental crust composition.

The neutron flux spectra in the crust from spontaneous fission and ($\alpha,n$) neutrons are shown in the Figure \ref{Figure: Neutron_flux}. For comparison, the cosmic-ray muon-induced neutron flux spectrum at 500 mwe is also plotted in the same figure. 
Reported in Table \ref{table:Radiogenic neutron flux} is the total neutron flux in the crust from each of the considered radiogenic neutron sources. The results show that the cosmic-ray muon-induced flux, even at 500 mwe depth in the crust, is over two orders of magnitude smaller than the assumed radiogenic neutron flux. However, it should be noted that the radiogenic neutron flux depends strongly on the uranium and thorium content in the rock, as well as on the composition of the rock.

 \begin{table}[h!]
 \small

\begin{tabular}{|l|c|}
 \hline
  Neutron Source & Total neutron flux\\
  & (neutrons/cm$^2$/sec)\\  
 \hline
 238U spont. fission & 3.21 $\times$ $10^{-5}$\\
 \hline
 238U ($\alpha$,n) & 9.92 $\times$ $10^{-5}$\\
 \hline
 235U ($\alpha$,n) & 4.56 $\times$ $10^{-8}$\\
 \hline
 232Th $\alpha$,n) & 1.91 $\times$ $10^{-4}$ \\
 \hline
 \end{tabular}
  \caption{Radiogenic neutron flux from spontaneous fission and ($\alpha,n$) neutrons from uranium and thorium decay chains in the crust.}
   \label{table:Radiogenic neutron flux}

\end{table}

\subsection{Radiogenic Production Rates}
\begin{figure}
\centering
 
\subfloat[]{
	\label{subfigures:a}
	\includegraphics[width=0.5\textwidth]{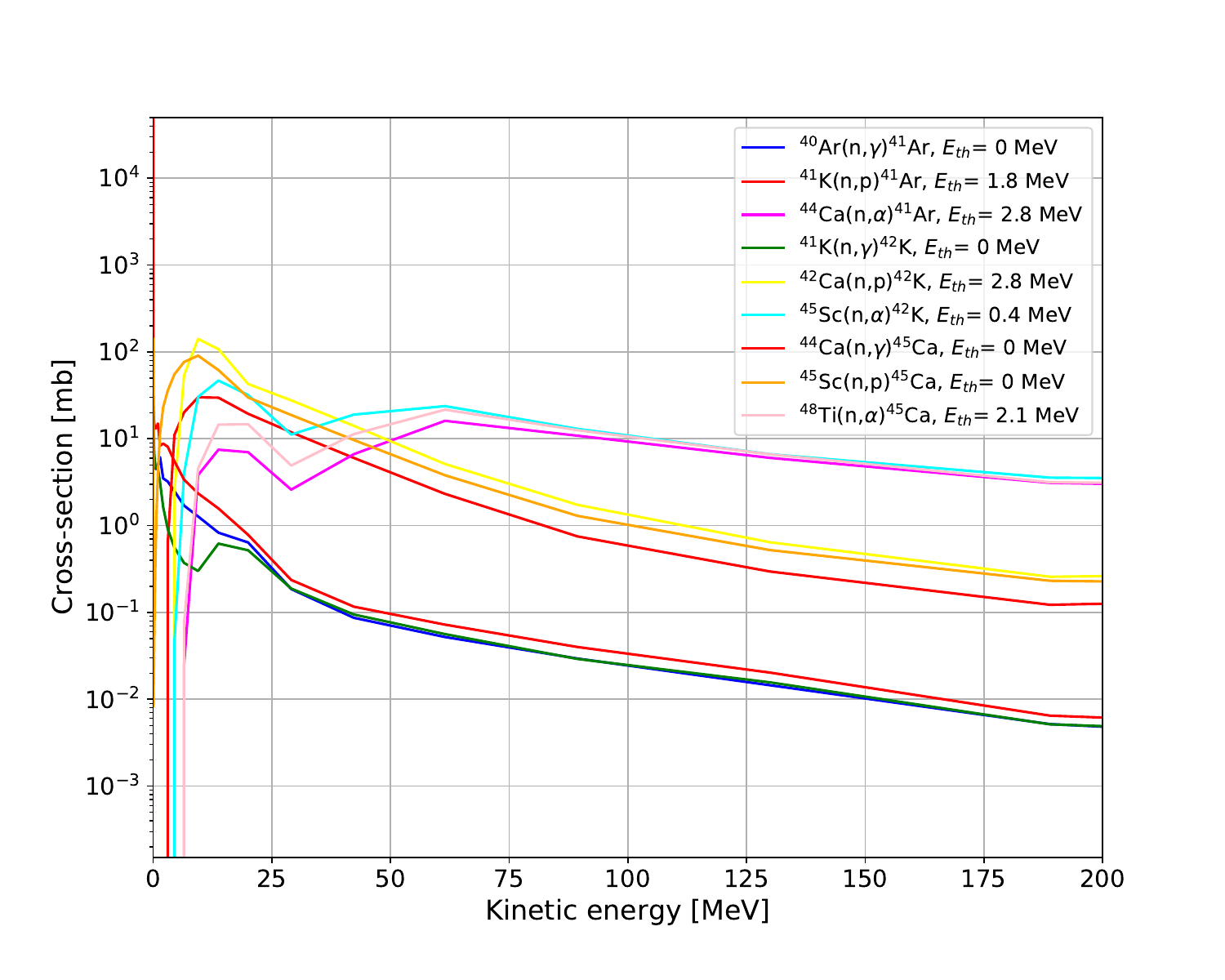} } 
 
\subfloat[]{
	\label{subfigures:b}
	\includegraphics[width=0.5\textwidth]{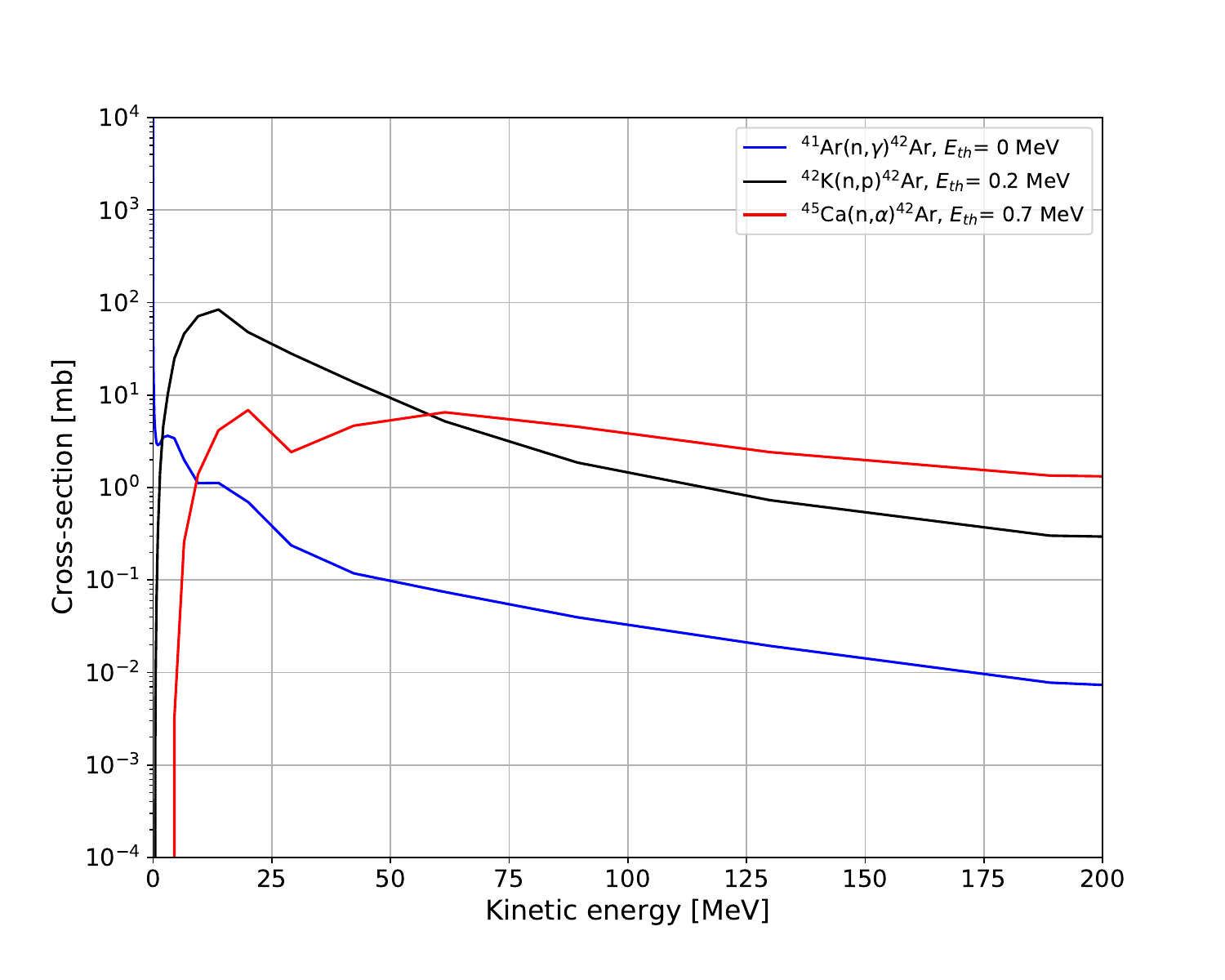}}

 \caption{(a)\isotope{Ar}{41}, \isotope{K}{42}, and \isotope{Ca}{45} production cross-sections for (n,$\gamma$), (n,p) and (n,$\alpha$) reactions. (b) \isotope{Ar}{42} production cross-sections (from TALYS) for reactions considered to estimate the radiogenic production of \isotope{Ar}{42}. Peak cross-section for \isotope{Ar}{41}(n,$\gamma$)\isotope{Ar}{42} is $1.42 \times 10^6$ millibarn at 0.07 eV. }
\label{fig:Intermediate_cross_sections} 
\end{figure}

Using Eq. (\ref{production_yield}) for selected (n,$\gamma$), (n,p) and (n,$\alpha$) reactions, we estimate the cosmogenic and radiogenic production rates of isotopes \isotope{Ar}{41}, \isotope{K}{42}, and \isotope{Ca}{45} in the Earth's crust. The TALYS-given cross-sections for the reactions considered are shown in Figure \ref{fig:Intermediate_cross_sections}a. $E_{th}$ represents the energy thresholds for the reactions. The production rates of the isotopes \isotope{Ar}{41}, \isotope{K}{42}, and \isotope{Ca}{45} are shown in the Table \ref{table:Intermediate_rates}. 
(n,$\gamma$) reactions are found to be the dominant production channels for those isotopes. 

\begin{table}[h!]
\small
\begin{tabular}{|p{1.1cm}|p{1.28cm}|p{1.79cm}|p{1.79cm}|p{2.1cm}|}
 \hline
  Isotopes & Isotope half-life & Cosmogenic \newline production rate (/ton/yr)\newline at 500 mwe & Radiogenic production \newline rate (/ton/yr) & Dominant production channel\\  
 \hline
 \isotope{Ar}{41} & 109 min & 39.6 & $1.82 \times 10^4$ & \isotope{Ar}{40}(n,$\gamma$)\isotope{Ar}{41}\\
 \hline
 \isotope{K}{42} & 12 hr & $4.89 \times 10^4$  & $1.56 \times 10^7$ & \isotope{K}{41}(n,$\gamma$)\isotope{K}{42}\\
 \hline
 \isotope{Ca}{45} & 163 d & $4.62 \times 10^4$  & $1.47 \times 10^7$ & \isotope{Ca}{44}(n,$\gamma$)\isotope{Ca}{45}\\
 \hline

 \end{tabular}
 \caption{\isotope{Ar}{41}, \isotope{K}{42}, \isotope{Ca}{45} production in the earth's crust obtained using TALYS for selected nuclear reactions.}
 \label{table:Intermediate_rates}
\end{table}

\begin{table}[h!]
\label{Table_radiogenic_production_rates}
 \small
\begin{tabular}{|p{2.1cm}|p{1.5cm}|p{2.0cm}|p{1.8cm}|}
 \hline
  Nuclear reaction & Reaction \newline threshold \newline (MeV) & Equilibrium \newline concentration of \isotope{Ar}{42} \newline (atoms/target atom) & Production rate \newline (/ton/yr)\\
  \hline
 \isotope{Ar}{41}(n,$\gamma$)\isotope{Ar}{42} & 0 & 4.13 $\times$ $10^{-17}$  & 4.75 $\times$ $10^{-18}$\\
 \hline
 \isotope{K}{42}(n,p)\isotope{Ar}{42} & 0 & $5.00 \times 10^{-23}$ & 3.35 $\times$ $10^{-20}$\\
 \hline
 \isotope{Ca}{45}(n,$\alpha$)\isotope{Ar}{42} & 0.7 & 1.84 $\times 10^{-26}$ & 3.68 $\times 10^{-21}$\\
 \hline
 \hline

 \end{tabular}
 \caption{Radiogenic \isotope{Ar}{42} production in the crust.
 Two-step reactions passing through intermediate radioactive isotopes \isotope{Ar}{41}, \isotope{K}{42}, and \isotope{Ca}{45} are only considered as direct production may not be possible for radiogenic particles considering the high-energy threshold of the reactions.}
 \label{table:radio_42Ar}
\end{table}

Radiogenic \isotope{Ar}{42} production is calculated by using Eq. \ref{production_yield} for neutron-induced reactions on  isotopes \isotope{Ar}{41}, \isotope{K}{42}, and \isotope{Ca}{45}. The equilibrium concentration of these isotopes in the crust were calculated from their production rates (shown in Table \ref{table:Intermediate_rates}) and used as an input for Eq. \ref{production_yield} to estimate the radiogenic \isotope{Ar}{42} production from nuclear reactions on these isotopes. The TALYS-given cross-sections for the relevant reactions are given in the Figure \ref{fig:Intermediate_cross_sections}b. 

Radiogenic production rates of \isotope{Ar}{42} are given in Table \ref{table:radio_42Ar}. The  production of \isotope{Ar}{42} for a crustal composition is estimated to be $4.79 \times 10^{-18}$ atoms per ton per year. The rate is many orders of magnitude smaller than the cosmogenic production rates obtained for 500 mwe and 3,000 mwe on the Earth's crust. From the table, one can observe the \isotope{Ar}{42} production rate is highest for the reaction \isotope{Ar}{41}(n,$\gamma$)\isotope{Ar}{42}, and by several orders of magnitude. The radiogenic production rate, however, is very sensitive to the concentration of \isotope{Ar}{40} assumed in the estimate of \isotope{Ar}{41} production. The assumed concentration of argon is 3 ppm \cite{rumble2020crc}.  The results show \isotope{Ar}{41}, \isotope{K}{42}, and \isotope{Ca}{45} are not produced enough to generate any significant level of \isotope{Ar}{42} radiogenically. A similar calculation for \isotope{Ar}{42} production from neutron-induced interaction on \isotope{K}{43} [$\tau_{1/2}$ = 22 hr] showed the \isotope{Ar}{42} production rate from that channel to be over 10 orders of magnitude smaller. 

As mentioned before, one-step radiogenic neutron-induced reactions are expected to have extremely low production rates because the radiogenic neutron energies are typically well below the reaction thresholds required to produce \isotope{Ar}{42}. Though not fully explored in this work, it may be possible to extract an upper limit on the \isotope{Ar}{42} production from direct one-step reactions by extrapolating the tail of the radiogenic neutron flux spectrum.   One relevant reaction would be \isotope{Ca}{43}(n,p)\isotope{Ar}{42}, for which  TALYS gives non-zero cross-sections above 15 MeV. With the sampling of the radiogenic neutron energies in our simulations, at 10 MeV, the radiogenic neutron flux is $1.6 \times 10^{-9}$ neutrons/MeV/cm$^2$/sec. However, the radiogenic neutron yield falls rapidly beyond 10 MeV (the ($\alpha$,n) neutron yield output of NeuCBOT shows the radiogenic neutron yield drops by 30 orders of magnitude between 10 MeV and 15 MeV. 

\section{\isotope{Ar}{42} activity in underground argon}
\label{sec:final_activity}
In the previous sections we have estimated the underground production rate of \isotope{Ar}{42} per unit mass of crustal rock. For dark matter and neutrino experiments that aim to utilize underground argon, the critical value of interest is the concentration of \isotope{Ar}{42} in argon extracted from deep underground sources such as the one used by DarkSide-50. Estimating the \isotope{Ar}{42} content in underground argon based on the production rate in rock is difficult as it involves estimating the diffusion of argon out of the rock, which can depend on many geological factors and is beyond the scope of this work. However, we expect \isotope{Ar}{42} and \isotope{Ar}{39} to have very similar diffusion constants and so we can use the ratio of the measured \isotope{Ar}{39} activity in underground argon to the calculated underground production rate of \isotope{Ar}{39} by Sramek et al. \cite{vsramek2017subterranean} to estimate the corresponding activity of \isotope{Ar}{42} in underground argon.

Below a few hundred meters-water-equivalent depth in the crust, \isotope{Ar}{39} production is primarily from \isotope{K}{39}(n,p)\isotope{Ar}{39} reactions induced by radiogenic neutrons \cite{Mei_underground_PhysRevC.81.055802,vsramek2017subterranean}. Using very similar calculations to the ones presented in this paper, Sramek et al. \cite{vsramek2017subterranean} estimated the \isotope{Ar}{39} production rate in K-Th-U-rich upper continental crust to be $2.9 \times 10^4$ \isotope{Ar}{39} atoms per ton of crust per year. We therefore have the following set of measurements and estimates for the production of argon radioisotopes in underground argon:
\begin{itemize}
    \item Estimated \isotope{Ar}{39} production rates in K-Th-U-rich upper continental crust \cite{vsramek2017subterranean}:  $2.9 \times 10^4$ \isotope{Ar}{39} atoms per ton of crust per year.
    \item Measured \isotope{Ar}{39} activity in underground argon \cite{darkside_UAr_DS50_39Ar}: $7.3 \times 10^{-4}$ Bq per kg of argon which corresponds to a steady-state production rate of $2.3 \times 10^{7}$ \isotope{Ar}{39} atoms per ton of argon per year.
    \item Estimated \isotope{Ar}{42} production rate in the Earth's crust at 3,000 mwe depth (from this work): $1.8 \times 10^{-3}$ \isotope{Ar}{42} atoms per ton of crust per year.
\end{itemize}
If one assumes the same ratio of concentrations for \isotope{Ar}{42} in the crustal rock to that in underground argon as in the above measurements and estimates for \isotope{Ar}{39}, then one would predict 1.4 \isotope{Ar}{42} atoms  per ton of argon per year at 3,000 mwe. Given this production rate, at equilibrium, \isotope{Ar}{42}-specific radioactivity in argon is estimated to be 1.4 decays per ton of argon per year.   A summary of the \isotope{Ar}{42} and \isotope{Ar}{39} production rates in crust and activities in argon are shown in the Table \ref{tab:results_summary}. 

The above estimate is expected to be conservative for a number of reasons: 
\begin{itemize}
   \item The percentage of \isotope{Ar}{42} atoms diffusing out of the rock and getting collected in the underground gas fields is likely smaller, given the shorter half-life of \isotope{Ar}{42} compared to \isotope{Ar}{39} (factor of 8).  
    \item The Doe Canyon gas wells (9,000 feet, 2.7 km depth) \cite{Doenichols2014subsurface} from where DarkSide extracted the underground argon are much deeper than 3,000 mwe. Argon in the gas wells likely originated in the mantle where the \isotope{Ar}{42} production is even more suppressed relative to \isotope{Ar}{39} (due to the primary \isotope{Ar}{42}-production mechanism being cosmogenic).
    \item The true \isotope{Ar}{39} activity in the underground argon could be significantly smaller, with the measured value likely due to an air incursion \cite{incursion_renshaw1239080procuring}.
\end{itemize}

\section{\isotope{Ar}{42} in isolated argon-containing gas pocket}
\begin{figure}[h]
\centering
\includegraphics[width=1.0\linewidth]{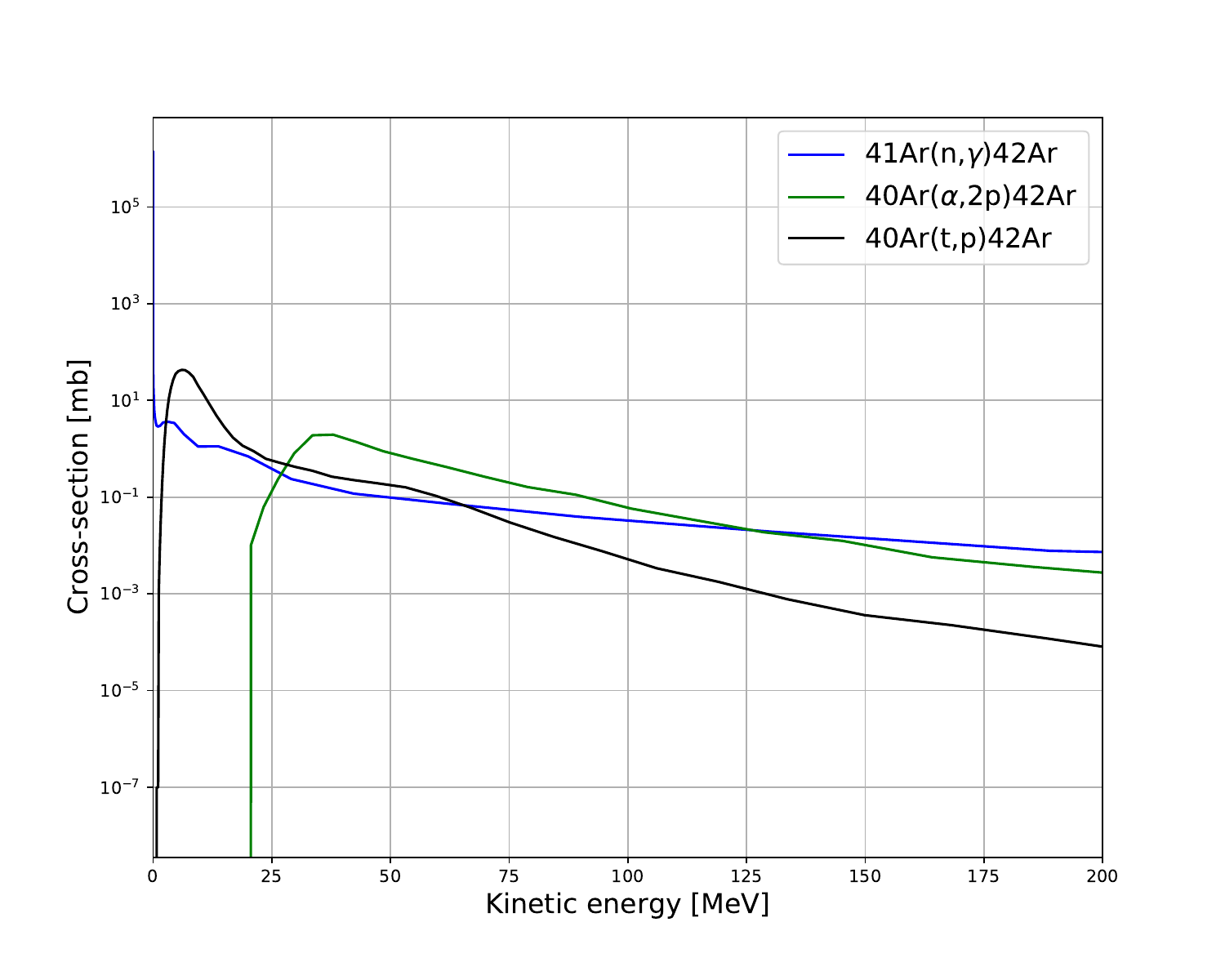}
\caption{\isotope{Ar}{42} production cross-sections as a function of kinetic energy.}
\label{Figure: 42Ar_from_40Ar}
\end{figure}

In addition to considering production of \isotope{Ar}{42} in crustal rock, we also considered production on argon gas that may be trapped deep underground.  Assuming the particle flux in the isolated argon gas pocket is the same as in the crust, we calculate the equilibrium ratio of \isotope{Ar}{42}/\isotope{Ar}{40} using Eq. (\ref{production_yield}) for selected reactions.
The relevant reactions for \isotope{Ar}{42} production in argon are i) \isotope{Ar}{40}(n,$\gamma$)\isotope{Ar}{41}, \isotope{Ar}{41}(n,$\gamma$)\isotope{Ar}{42}; ii) \isotope{Ar}{40}($\alpha$,2p)\isotope{Ar}{42}; and iii) \isotope{Ar}{40}(t,p)\isotope{Ar}{42}. The TALYS cross-sections for the reactions are shown in the Figure \ref{Figure: 42Ar_from_40Ar}.

  In argon, \isotope{Ar}{40}(t,p)\isotope{Ar}{42} is found to be the dominant production channel, with contribution from \isotope{Ar}{40}($\alpha$,2p)\isotope{Ar}{42} an order of magnitude smaller. This may not appear surprising given the smaller threshold of the \isotope{Ar}{40}(t,p)\isotope{Ar}{42} reaction and cosmic-ray muon-induced triton flux comparable to the $>$ 10 MeV $\alpha$ flux. Radiogenic production through \isotope{Ar}{41}(n,$\gamma$)\isotope{Ar}{42} is insignificant compared to other channels mentioned above. For large gas pockets or gas fields, which may even have other gas species, particle (particularly triton and alpha) production has to be simulated in the rock as well as in the modeled gas fields. 
  
  We obtain an \isotope{Ar}{42}/\isotope{Ar}{40} ratio of $4.64 \times 10^{-28}$ ($8.38 \times 10^{-32}$) at 500 mwe (3,000 mwe) depth in the crust. This is equivalent to  0.147 (1.09 $\times$ 10$^{-3})$ \isotope{Ar}{42} atoms per ton of argon at 500 mwe (3,000 mwe) depth. 
  This estimated rate of \isotope{Ar}{42} production from \isotope{Ar}{40} is roughly three orders of magnitude smaller than the estimate from production on crustal rock given in Section \ref{sec:final_activity}. 
  However, above rate does not include the contribution from argon produced in the rock that potentially can diffuse out and and collected in the argon reservoir.


\section{Uncertainties}
\label{sec: uncertainties}
\begin{figure}[h]
\centering
\includegraphics[width=1.0\linewidth]{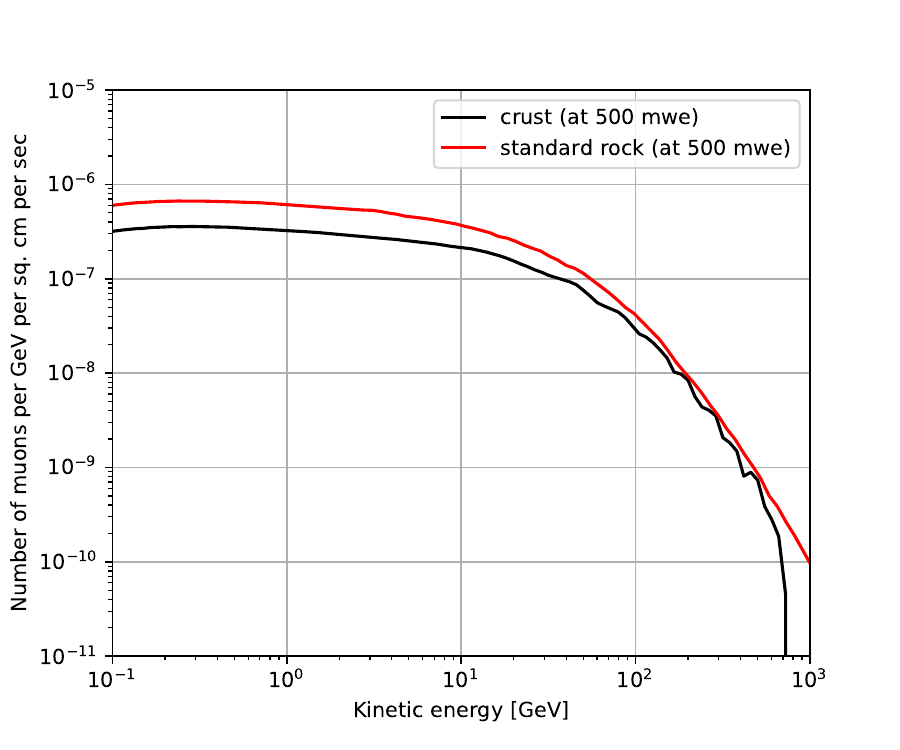}
\caption{Differential muon flux spectrum at 500 mwe in the crust. Cosmic-ray muon spectrum obtained by propagating the EXPACS-given overground (location: Doe Canyon, CO) muons through 500 mwe in the modeled crust (in black). Cosmic-ray muon spectrum at 500 mwe for MUSIC-given input muon spectrum for a standard rock at 500 mwe (in red). Total muon flux and mean muon energies shown in the Table \ref{table:muon flux comparison 500 mwe}. Note: All muons were propagated vertically.}
\label{Figure: Muon_flux_comparison}

\end{figure}

The total production rates reported in the paper have a  statistical uncertainty of roughly $10\%$, with systematic uncertainties expected to dominate. Our results show that \isotope{Ar}{42} production is primarily cosmogenic. The \isotope{Ar}{42} production rates are expected to be correct within a factor of 3 to 5. Major uncertainty is expected to be from systematics associated with hadronic processes. Use of MUSIC-code-generated muon spectra and flux for a standard rock for our crustal composition is expected to introduce a systematic uncertainty of $\sim$ 60-100$\%$ based on the rock composition dependence on the total muon flux reported in \cite{lechmann2018effect} and our comparison of differences in muon flux for a standard rock versus continental crust shown in Figure \ref{Figure: Muon_flux_comparison}.  Since all muons were propagated vertically, we have underestimated the particle and isotope production yield. Assuming the cosmic-ray muon-induced secondary particle yield and isotope production yield dependence to mean muon energy follows the simple parameterisation,  $\bar{E}^{-0.7}_{\mu}$ reported in \cite{ryajskaya1965depth}, the yields are underestimated by $<$ 20$\%$.  Since particle interactions on isotopes of calcium, mainly \isotope{Ca}{44} and \isotope{Ca}{48}, contribute significantly to \isotope{Ar}{42} production,  any significant presence of a calcium-rich mineral like limestone in the assumed rock composition can drive the production rates by a few factors.

The \isotope{Ar}{42} production rates at a given depth, to a first order, can be calculated by scaling the rates obtained in this work by a factor x $\bar{E}^{-0.7}_{\mu}$ /$\mu_{flux}$, with $\mu_{flux}$ and $\bar{E}_{\mu}$ for different depths for a standard rock \cite{KUDRYAVTSEV2009339}. However, at larger depths, the cosmic-ray muon energy losses and muon flux attenuation becomes increasingly composition dependent, systematics associated with the muon flux get larger, and the production rates obtained with this scaling are expected to be correct only in an order of magnitude.

Our study suggests that the \isotope{Ar}{42} production rate decreases with depth, as its production is primarily cosmogenic. However, at very large crustal/mantle depths, it is likely that the production of \isotope{Ar}{42} is primarily due to the interaction of muons that are generated by neutrinos \cite{crouch1978cosmic}.
\section{Results and Discussion}
\label{sec:discussion}
We have estimated the \isotope{Ar}{42} production rates at 500 mwe and 3,000 mwe depth in the Earth's crust. We find that radiogenic production is insignificant and that cosmogenic production is expected to dominate up to large crustal depths. At a depth of 3,000 mwe, the expected  \isotope{Ar}{42} production rate is 1.8 $\times$ $10^{-3}$ atoms per ton of the crust per year, seven orders of magnitude smaller than the \isotope{Ar}{39} production rate calculated in  \cite{vsramek2017subterranean,Mei_underground_PhysRevC.81.055802}. 

\begin{table}[h!]
\label{Cosmogenic_production}
 \small
\begin{tabular}{|p{2.1cm}|p{1.9cm}|p{1.9cm}|p{2.4cm}| }
 \hline
 Reactions & TALYS-based production rate from selected reactions [atoms/ton/y]& FLUKA residual-nuclei-recording-based)\newline[atoms/ton/y]  & Major \isotope{Ar}{42} yielding reactions \\  
 \hline
 n,p,$\alpha$,d,t- \newline induced reactions &  $2.5 \times 10^{-4}$ & $4.2 \times 10^{-4}$& \isotope{Ca}{44}($n$,3He)\isotope{Ar}{42}\newline \\
 \hline
 Heavy-ion collisions  & -- & $8.3 \times 10^{-4}$ & \isotope{Fe}{56}(H*,X)\isotope{Ar}{42} \newline \isotope{Ca}{44}(H*,X)\isotope{Ar}{42} \newline \isotope{Ca}{48}(H*,X)\isotope{Ar}{42}\\
 \hline
 Photon-induced reactions & -- & $1.6 \times 10^{-4}$ & \isotope{Ca}{44}($\gamma$,X)\isotope{Ar}{42} \\
 \hline
 Pion-induced reactions & -- & $1.6 \times 10^{-4}$ &  \isotope{Fe}{56}($\Pi^-$,X)\\
 \hline
 Other \newline cosmic-ray muon-induced reactions &-- & $2.0 \times 10^{-4}$& \isotope{Ca}{44}($\mu^-$,2p)\isotope{Ar}{42} \newline \isotope{Cl}{42} $\beta^-$ decay\\
 \hline
 \hline
 \hline
 Radiogenic reactions & $4.8 \times 10^{-18}$ & --& \isotope{Ar}{41}(n,$\gamma$)\isotope{Ar}{42}\\ 
 \hline
 \hline
 All reactions (sum) & $2.5 \times 10^{-4}$ & {$1.8 \times 10^{-3}$} & --\\
 \hline
 \end{tabular}
 \caption{\isotope{Ar}{42} production rates at 3,000 mwe in the crust (rates are given in atoms per ton of crust per year). Cosmogenic production is expected to dominate the radiogenic production up to significant depths in the crust.}
 \label{table:production-table-final}
\end{table}


\begin{table}[]
    \centering
    \begin{tabular}{|l|c|c| }
        \hline
         Isotope & Production rate in crust & Specific radioactivity in argon\\
          & (atoms/ton (rock)/yr) & (decays/ton (argon)/yr)\\
         \hline
         \isotope{Ar}{39} & $2.9 \times 10^4$ \cite{vsramek2017subterranean} & $2.3 \times 10^7$ \cite{darkside_UAr_DS50_39Ar} \\
         \hline
         \isotope{Ar}{42} & $1.8 \times 10^{-3}$ & 1.4 \\
         \hline
    \end{tabular}
    \caption{Estimated production rates of \isotope{Ar}{42} at 3,000 mwe compared to \isotope{Ar}{39}. The \isotope{Ar}{42} activity in gas from atoms produced in the rock is estimated by scaling the measured \isotope{Ar}{39} activity in gas by the ratio of calculated production rates in rock. See discussion in Section \ref{sec:final_activity}.
    \label{tab:results_summary}}
\end{table}
The activity of \isotope{Ar}{42} in underground argon from the Doe Canyon wells was estimated using the \isotope{Ar}{39} activity measured by DarkSide and calculations of the \isotope{Ar}{39} production in crustal rock, under the assumption that diffusion of \isotope{Ar}{42} out of the rock is similar to \isotope{Ar}{39}.  The \isotope{Ar}{42} activity in underground argon extracted is estimated to be 1.4 decays per ton of argon per year. For reasons discussed in Section \ref{sec:final_activity}, this is expected to be an upper limit.

Based on the estimates reported in this paper, argon extracted from underground sources is expected to be significantly depleted of \isotope{Ar}{42}, with a larger depletion factor than \isotope{Ar}{39}. Use of underground argon will greatly enhance the physics capabilities of kton-scale argon-based neutrino measurements like the ones proposed in \cite{low_caratelli2022low,low_avasthi2022low, bezerra2023large}. The LEGEND experiment \cite{legend_abgrall2021legend} could also greatly benefit from the use of underground argon. With atmospheric argon, the background rate from \isotope{Ar}{42}/\isotope{K}{42} (before analysis cuts) is expected to be 0.72 cts/kg/yr/keV \cite{legend_abgrall2021legend}. With the use of underground argon, as per our estimate, the \isotope{Ar}{42}/\isotope{K}{42} background suppression could be a factor of $10^7$ or higher. More concerning for large-scale argon-based experiments could be a possible infiltration of atmospheric argon or/and cosmogenic production of \isotope{Ar}{42} in the extracted underground argon during storage, transport, or extended cosmic-ray exposure above ground \cite{ZHANG2022102733}.

\section{Acknowledgements}
We would like to extend our thanks to Prof. Vitaly A. Kudryavtsev for the insightful discussions we had on the MUSIC code and muon propagation. Additionally, we extend our appreciation to Prof. Shawn Westerdale for his guidance on the NeuCBOT code and for engaging discussions regarding radiogenic ($\alpha,n$) neutron yields.
This work was funded by Laboratory Directed Research and Development (LDRD) at Pacific Northwest National Laboratory (PNNL). PNNL is operated by Battelle Memorial Institute for the U.S. Department of Energy (DOE) under Contract No. DE-AC05-76RL01830. Parts of this study at PNNL were supported by the DOE, USA Office of High Energy Physics Advanced Technology R\&D subprogram.
\bibliography{apssamp}
\section{Appendix}
\subsection{List of nuclear reaction considered in  TALYS-based \isotope{Ar}{42} production estimate}
The list of the nuclear reactions considered in the TALYS-based \isotope{Ar}{42} production estimate and energy thresholds for those reactions are listed in Table \ref{table:list of reactions}.
The target isotope list includes stable and some long-lived isotopes in ($Z_{Ar}-4, Z_{Ar}+4$) (Z: atomic number of an element), and also  Fe and Mn isotopes. Latter isotopes are also considered because of their relatively high abundance in the Earth's crust.
The target isotopes and information about the nuclear reaction is also provided in Table \ref{table:list of reactions}).

\begin{table}
\small

\caption{Isotopes and the reactions considered for TALYS-based estimate of \isotope{Ar}{42} production. Only reactions induced by neutron, proton, deuteron, and triton were considered.}
\begin{tabular}{|p{1.1cm}|p{3.4cm}|p{2.5cm}|p{1.0cm}| }
 \hline
 \hline
Element \newline (Z)  &  Isotopes \newline (Isotopic abundance) & Reactions & Energy threshold (MeV)\\  
 \hline
 Si (Z$=$14) & \isotope{Si}{28}(92.23$\%$) \isotope{Si}{29}(4.67$\%$) \isotope{Si}{30}(3.10$\%$) & - & - \\
 \hline
 \hline
 P (Z$=$15) & \isotope{P}{31}(100$\%$ & - & - \\
 \hline
 \hline
 S (Z$=$16)& \isotope{S}{32}(95.02\%) \isotope{S}{33}(0.75\%) \newline\isotope{S}{34}(4.21\%) \newline \isotope{S}{36}(0.02\%) \newline \isotope{S}{38}(trace, $\tau_{1/2}$=3h)& \isotope{S}{38}($\alpha$,$\gamma$)\isotope{Ar}{42} \newline   & 0.0 \\
 \hline
 \hline
 Cl (Z$=$17) & \isotope{Cl}{35}(75.77\%) \isotope{Cl}{37}(24.23\%) \isotope{Cl}{42}(trace,$\tau_{1/2}$=6s)  \isotope{Cl}{43}(trace,$\tau_{1/2}$=280ms) & \isotope{Cl}{42} $\beta^-$decay \newline \isotope{Cl}{43} $\beta^-$+n decay & -\\ 
 \hline
 \hline
 Ar (Z$=$18) & \isotope{Ar}{36}(0.337\%) \isotope{Ar}{38}(0.063\%) \isotope{Ar}{40}(99.6\%) \isotope{Ar}{41}(trace,$\tau_{1/2}$=109m) &  \isotope{Ar}{40}($\alpha$,2p)\isotope{Ar}{42} \isotope{Ar}{41}(n,$\gamma$)\isotope{Ar}{42}  \isotope{Ar}{40}(t,p)\isotope{Ar}{42} & 14.0 \newline 0 \newline 0.7 \\
 \hline
 \hline
 K (Z$=$19) & \isotope{K}{39}(93.3\%) \isotope{K}{40}(0.011\%) \isotope{K}{41}(6.73\%) \isotope{K}{42}(trace,$\tau_{1/2}$=12h) \isotope{K}{43}(trace,$\tau_{1/2}$=22h) & \isotope{K}{41}($\alpha$,X)\isotope{Ar}{42} \isotope{K}{41}(t,2p)\isotope{Ar}{42} \isotope{K}{42}(n,p)\isotope{Ar}{42} \isotope{K}{43}(n,d)\isotope{Ar}{42} & 22.6 \newline 0.8 \newline 0.2 \newline 7.4 \\
 \hline
 \hline
 Ca (Z$=$20) & \isotope{Ca}{40}(96.9\%) \isotope{Ca}{42}(0.647\%) \isotope{Ca}{43}(0.135\%) \isotope{Ca}{44}(2.086\%) \isotope{Ca}{45}(trace,$\tau_{1/2}$=163d)  \isotope{Ca}{46}(0.004\%) \newline \isotope{Ca}{48}(0.187\%) & \isotope{Ca}{42}($\alpha$,X)\isotope{Ar}{42} \isotope{Ca}{43}($\alpha$,X)\isotope{Ar}{42} \isotope{Ca}{44}($\alpha$,X)\isotope{Ar}{42} \isotope{Ca}{45}($\alpha$,X)\isotope{Ar}{42} \isotope{Ca}{46}($\alpha$,X)\isotope{Ar}{42} \isotope{Ca}{48}($\alpha$,X)\isotope{Ar}{42} \isotope{Ca}{43}(n,2p)\isotope{Ar}{42} \isotope{Ca}{44}(n,3He)\isotope{Ar}{42} \isotope{Ca}{45}(n,$\alpha$)\isotope{Ar}{42} \isotope{Ca}{46}(n,$\alpha$+n)\isotope{Ar}{42} \isotope{Ca}{48}(n,X)\isotope{Ar}{42} \isotope{Ca}{44}(p,3p)\isotope{Ar}{42} \isotope{Ca}{43}(d,2p)\isotope{Ar}{42} \isotope{Ca}{44}(d,X)\isotope{Ar}{42} \isotope{Ca}{46}(d,X)\isotope{Ar}{42} \isotope{Ca}{48}(d,X)\isotope{Ar}{42}  & 33.8 34.0 25.1 27.6 12.1 21.6 10.7 14.2 \newline 0.7 \newline 11.4 28.9 22.1 13.1 16.9 11.6 18.3\\
 \hline
 \hline
 Sc (Z$=$21) & \isotope{Sc}{45}(100\%) \isotope{Sc}{46}(trace,$\tau_{1/2}$=84d) & \isotope{Sc}{45}(n,X)\isotope{Ar}{42} \isotope{Sc}{45}(p,4p)\isotope{Ar}{42} \isotope{Sc}{45}(d,X)\isotope{Ar}{42} &21.3 \newline 29.1 \newline 24.0\\
 \hline
 \hline
 \label{table:list of reactions}
 \end{tabular}

 \end{table}

 \begin{table}
\small
 \begin{tabular}{|p{1.1cm}|p{3.6cm}|p{2.4cm}|p{1.0cm}| }
 \hline
 \hline
Element \newline (Z)  &  Isotopes \newline (Isotopic abundance) & Reactions  & Energy threshold (MeV)\\  
 \hline
 
 Ti (Z$=$22) & \isotope{Ti}{46}(8.0\%) \newline \isotope{Ti}{47}(7.3\%) \newline \isotope{Ti}{48}(73.8\%) \newline \isotope{Ti}{49}(5.5\%) \newline \isotope{Ti}{50}(5.4\%) & \isotope{Ti}{46}($\alpha$,X)\isotope{Ar}{42} \isotope{Ti}{47}($\alpha$,X)\isotope{Ar}{42} \isotope{Ti}{48}($\alpha$,X)\isotope{Ar}{42} \isotope{Ti}{49}($\alpha$,X)\isotope{Ar}{42} \isotope{Ti}{50}($\alpha$,X)\isotope{Ar}{42} \isotope{Ti}{46}(n,X)\isotope{Ar}{42} \isotope{Ti}{47}(n,X)\isotope{Ar}{42} \isotope{Ti}{48}(n,X)\isotope{Ar}{42} \isotope{Ti}{49}(n,X)\isotope{Ar}{42} \isotope{Ti}{50}(n,X)\isotope{Ar}{42} \isotope{Ti}{46}(p,X)\isotope{Ar}{42} \isotope{Ti}{47}(p,X)\isotope{Ar}{42} \isotope{Ti}{48}(p,X)\isotope{Ar}{42} \isotope{Ti}{49}(p,X)\isotope{Ar}{42} \isotope{Ti}{50}(p,X)\isotope{Ar}{42} \isotope{Ti}{46}(d,X)\isotope{Ar}{42} \isotope{Ti}{47}(d,X)\isotope{Ar}{42} \isotope{Ti}{48}(d,X)\isotope{Ar}{42} \isotope{Ti}{49}(d,X)\isotope{Ar}{42} \isotope{Ti}{50}(d,X)\isotope{Ar}{42} \isotope{Ti}{46}(t,X)\isotope{Ar}{42} \isotope{Ti}{47}(t,X)\isotope{Ar}{42} \isotope{Ti}{48}(t,X)\isotope{Ar}{42} \isotope{Ti}{49}(t,X)\isotope{Ar}{42} \isotope{Ti}{50}(t,X)\isotope{Ar}{42} & 46.0 40.2 29.6 24.2 15.8 31.8 19.9 23.8 11.1 22.3 39.7 40.8 31.7 32.1 22.3 34.8 22.6 24.9 13.7 16.6 20.3 21.5 12.0 11.9 11.4 \\
 \hline
 \hline
 Mn (Z$=$25) & \isotope{Mn}{55}(100\%) \newline \isotope{Mn}{53}(trace,$\tau_{1/2}=$3.7E6y) &  \isotope{Mn}{55}($\alpha$,X)\isotope{Ar}{42} \isotope{Mn}{55}(n,X)\isotope{Ar}{42} \isotope{Mn}{55}(p,X)\isotope{Ar}{42} \isotope{Mn}{55}(d,X)\isotope{Ar}{42} \isotope{Mn}{55}(t,X)\isotope{Ar}{42} & 24.5 \newline 36.3 \newline 38.5 \newline 33.6 \newline 19.1 \\
 \hline
 \hline
 Fe (Z$=$26) & \isotope{Fe}{54}(5.85\%) \newline \isotope{Fe}{56}(91.75\%) \newline \isotope{Fe}{57}(2.12\%) \newline \isotope{Fe}{58}(0.28\%) \newline \isotope{Fe}{60}(trace,$\tau_{1/2}=$2.6E6y)& \isotope{Fe}{54}($\alpha$,X)\isotope{Ar}{42} \isotope{Fe}{54}(n,X)\isotope{Ar}{42} \isotope{Fe}{54}(p,X)\isotope{Ar}{42} \isotope{Fe}{54}(d,X)\isotope{Ar}{42} \isotope{Fe}{54}(t,X)\isotope{Ar}{42}
  \isotope{Fe}{56}($\alpha$,X)\isotope{Ar}{42} \isotope{Fe}{56}(n,X)\isotope{Ar}{42} \isotope{Fe}{56}(p,X)\isotope{Ar}{42} \isotope{Fe}{56}(d,X)\isotope{Ar}{42} \isotope{Fe}{56}(t,X)\isotope{Ar}{42} \isotope{Fe}{57}($\alpha$,X)\isotope{Ar}{42} \isotope{Fe}{57}(n,X)\isotope{Ar}{42} \isotope{Fe}{57}(p,X)\isotope{Ar}{42} \isotope{Fe}{57}(d,X)\isotope{Ar}{42} \isotope{Fe}{57}(t,X)\isotope{Ar}{42} \isotope{Fe}{58}($\alpha$,X)\isotope{Ar}{42} \isotope{Fe}{58}(n,X)\isotope{Ar}{42} \isotope{Fe}{58}(p,X)\isotope{Ar}{42} \isotope{Fe}{58}(d,X)\isotope{Ar}{42} \isotope{Fe}{58}(t,X)\isotope{Ar}{42} & 60.0  \newline  49.0 \newline  56.8 \newline  52.1 \newline  39.4  \newline  51.5 \newline  41.0  \newline  48.9 \newline  44.1 \newline  29.7 \newline  51.3 \newline  27.8  \newline  48.8 \newline  30.7 \newline  35.4  \newline  40.0 \newline  38.1 \newline  38.1 \newline  52.1 \newline  39.4\\
  \hline
  \hline
 \end{tabular}
 \end{table}

\subsection{Comparison of MUSIC versus EXPACS input muon spectra }
\begin{table}[h!]
 \caption{Total muon flux and mean muon energy at ~ 500 mwe for a standard rock and the modeled crust. Also, shown in the table is the Muon flux obtained by using FLUKA-USRTRACK tracking and event-by-event tracking using the FLUKA user routine.}
 
\begin{tabular}{cccc}
 \hline
 Input muon spectra & FLUKA Algorithm & Muon flux  & Mean energy \\  
 & & ($\mu$/cm$^2$/sec) & (GeV)\\
 \hline
 MUSIC & - & $1.94\times10^{-5}$ & 69\\
 \hline
\multirow{3}{*}{EXPACS} &  FLUKA-mgdraw & $1.5\times10^{-5}$ & \multirow{3}{*}{63} \\ 
& FLUKA-USRTRACK  & $1.3\times10^{-5}$ &  \\
 \hline
 \end{tabular}
 \label{table:muon flux comparison 500 mwe}
\end{table}
 
\end{document}